%% file: main.tex
\documentclass{article}

\usepackage{arxiv}

\usepackage[utf8]{inputenc} 
\usepackage[T1]{fontenc}    
\usepackage{hyperref}       
\usepackage{url}            
\usepackage{booktabs}       
\usepackage{amsfonts}       
\usepackage{amsmath}
\usepackage{nicefrac}       
\usepackage{microtype}      
\usepackage{graphicx}
\usepackage{natbib}
\usepackage{doi}
\usepackage[
  round-mode = places,
  round-precision = 2,
  detect-weight=true,
  round-pad = false 
]{siunitx}
\usepackage{cleveref}
\usepackage{subcaption}
\usepackage{tikz}
\usetikzlibrary{shapes,arrows,positioning,calc,backgrounds}

\def\tens#1{\ensuremath{\boldsymbol{\mathsf{#1}}}} 
\def\vec#1{\ensuremath{\boldsymbol{\mathbf{#1}}}}

\title{Adapting Deep Variational Bayes Filter for Enhanced Confidence Estimation in Finite Element Method Integrated Networks (FEMIN)}

\usepackage{authblk}

\newbox{\orcid}\sbox{\orcid}{\includegraphics[scale=0.06]{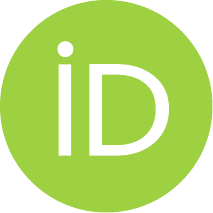}} 
\author[1, 2]{%
	\href{https://orcid.org/0000-0002-8410-6819}{\usebox{\orcid}\hspace{1mm}Simon Thel\thanks{\texttt{simon.thel@volkswagen.de}}}%
}
\author[1]{%
	\href{https://orcid.org/0000-0001-7763-6726}{\usebox{\orcid}\hspace{1mm}Lars Greve}%
}
\author[1]{%
	\href{https://orcid.org/0000-0001-8959-368X}{\usebox{\orcid}\hspace{1mm} Maximilian Karl}}
\author[1,3,4]{%
	\href{https://orcid.org/0000-0003-4418-4916}{\usebox{\orcid}\hspace{1mm} Patrick van der Smagt}}
\affil[1]{Volkswagen AG, Berliner Ring 2, 38440 Wolfsburg, Germany}
\affil[2]{TU Munich, School of Computation, Information and Technology, Arcisstraße 2, 80333 Munich,Germany}
\affil[3]{Department of Computer Science, ELTE University, Budapest, Hungary}
\affil[4]{Graduate School of Systemic Neurosciences, LMU, Munich, Germany}

\renewcommand{\shorttitle}{DVBF for FEMIN}

\hypersetup{
pdftitle={DVBF for FEMIN},
pdfsubject={q-bio.NC, q-bio.QM},
pdfauthor={Simon Thel},
pdfkeywords={Finite element method, crash simulation, Neural network, Probabilistic learning},
}

\begin{document}

\input{text}

\bibliographystyle{elsarticle-num-names}
\bibliography{PhD}  

\end{document}

%% file: text.tex
\maketitle
\begin{abstract}
The Finite Element Method (FEM) is a widely used technique for simulating crash scenarios with high accuracy and reliability.
To reduce the significant computational costs associated with FEM, the Finite Element Method Integrated Networks (FEMIN) framework integrates neural networks (NNs) with FEM solvers.
However, this integration can introduce errors and deviations from full-FEM simulations, highlighting the need for an additional metric to assess prediction confidence, especially when no ground truth data is available.
In this study, we adapt the Deep Variational Bayes Filter (DVBF) to the FEMIN framework, incorporating a probabilistic approach to provide qualitative insights into prediction confidence during FEMIN simulations.
The adaptation involves using the learned transition model for a predictive decoding step, generating a preliminary force prediction.
This predictive force is used alongside the displacement and the velocity data from the FEM solver as input for the encoder model.
The decoder reconstructs the likelihood distribution based on the posterior.
The mean force of this distribution is applied to the FEM solver, while the predicted standard deviation can be used for uncertainty estimation. 
Our findings demonstrate that the DVBF outperforms deterministic NN architectures in terms of accuracy.
Furthermore, the standard deviation derived from the decoder serves as a valuable qualitative metric for assessing the confidence in FEMIN simulations. 
This approach enhances the robustness of FEMIN by providing a measure of reliability alongside the simulation results.
\end{abstract}
\keywords{Finite element method \and Crash simulation \and Neural network \and Probabilistic learning}

\section{Introduction}

Crash simulations have become an indispensable tool in automotive engineering, crucial in vehicle safety evaluation and improvement \citep{klug2023euro}. 
These simulations, traditionally based on the Finite Element Method (FEM), provide high-fidelity simulation results to meet stringent safety standards \citep{nhtsa_2022}.
However, the complexity of these models results in substantial computational costs, with simulations often taking hours or even days to complete, despite advances in parallel computing \citep{law_parallel_1986, mafi_gpubased_2014,xue_jax-fem_2023}.

Our previous work introduced the Finite Element Method Integrated Networks (FEMIN) \citep{thel_introducing_2024}, a novel framework that combines Neural Network (NN) models with FEM simulations. 
FEMIN aims to reduce computational time by replacing entire regions in the simulation model with NN surrogates during runtime. 
This approach is advantageous in scenarios where engineers focus on evaluating the crash performance of specific localized regions of a vehicle within a full-vehicle context.
FEMIN is based on the premise that the impact of design changes on mechanical behavior decreases as the distance from the modified part increases.
Then, the vehicle is divided into two main parts:
\begin{enumerate}
    \item A substructure close to the part undergoing design changes, where the influence of these changes is assumed to be significant.
    \item The remainder of the vehicle, where the influence of the design change is assumed to be minor.
\end{enumerate}
The latter part is replaced by a reactive surrogate model based on the FEMIN approach, reducing computational cost while still allowing for detailed FEM analysis of the area of interest.

While FEMIN has shown promise in accelerating crash simulations, the integration of NNs with FEM solvers introduces new challenges, particularly in terms of accuracy and reliability. 
One critical aspect that needs to be addressed is the assessment of prediction confidence, especially in scenarios where ground truth data is unavailable for validation.

This study presents an innovative adaptation of the Deep Variational Bayes Filter (DVBF) to the FEMIN framework.
This adaptation introduces a probabilistic approach to FEMIN, providing qualitative insights into prediction confidence during FEMIN simulations.
By incorporating DVBF, we aim to enhance the reliability and interpretability of FEMIN results, addressing one of the critical challenges in adopting machine learning techniques in safety-critical applications such as automotive crash simulations.

Our approach involves several key modifications to the standard DVBF architecture \citep{karl_deep_2017-1} to align it with the requirements of the FEMIN framework.
We use the learned transition model for a predictive decoding step, generating a preliminary force prediction. This predictive force in conjunction with the kinematics\footnote{In the context of this work, "kinematic" refers to the displacement and velocity of the FEM simulation. Depending on the dimensionality of the simulation, the kinematics can include translational and rotational components.} from the FEM solver serves as input for the encoder model.
The decoder then reconstructs the likelihood distribution based on the posterior, from which we derive the mean force applied to the FEM solver.
This paper expands on our previous work by:
\begin{enumerate}
    \item Detailing the adaptation of DVBF to the FEMIN framework, making the adapted DVBF a generative and simultaneously an autoregressive model.
    \item Comparing the performance of DVBF against deterministic NN architectures in terms of accuracy.
    \item Assessing the uncertainty of the decoder output as a qualitative metric to estimate the prediction confidence in FEMIN.
\end{enumerate}

By addressing the critical issue of confidence estimation, this study aims to further enhance the robustness and reliability of the FEMIN approach, potentially accelerating its adoption in automotive design and safety evaluation processes. The integration of probabilistic methods not only improves the accuracy of simulations but also provides valuable insights into the reliability of predictions.

\section{Related Work}

While the FEMIN approach \citep{thel_introducing_2024} directly integrates a NN into the FEM solver to replace entire regions of the mesh, other recent research focuses on integrating NN as a constitutive model \citet{logarzo_smart_2021, tancogne-dejean_recurrent_2021}. 
For example, \citet{pantidis_integrated_2023} embedded NNs within finite element stiffness functions to accelerate the damage computations. 
Their physics-informed NN (PINN) predicts both strain and its gradient on an element basis, which leads to an accelerated construction of the Jacobian matrix.
NNs can also be used as a computationally efficient surrogate for FEM simulations \citep{liang_deep_2018, kudela_recent_2022, deshpande_magnet_2024}.
\citet{gulakala_graph_2023} used a graph NN (GNN) that directly predicts the von Mises stress of a FEM mesh. 
The stress distribution was then used as input for a standard FEM post-processor.
The surrogate modeling approach by \citet{haghighat_physics-informed_2021} used PINNs to predict von Mises stress. 
Their approach could be used for other field variables, and the results demonstrated transfer learning capabilities. 
\citet{van_de_weg_neural_2021} predicted fracture of tensile specimens using an LSTM-based learning approach. 
They showed excellent data efficiency while maintaining high accuracy.
Besides replacing the entire FEM simulations, NNs can also be used to accelerate multi-scale FEM simulations \citep{Ghavamian2019, danoun_fe-lstm_2024} or used as an element-based surrogate within a FEM simulation \citep{ouyang_neural_2024, koeppe_intelligent_2020}. 

Surrogate models are also widely used in uncertainty quantification for solid mechanics \citep{rocas_nonintrusive_2021, wojtkiewicz_uncertainty_2001}
In this context, probabilistic NN can be beneficial:
\citet{deshpande_probabilistic_2022} applied a Variational Bayes approach \citep{kingma_auto-encoding_2014} to the U-Net architecture \cite{ronneberger_u-net_2015}.
The probabilistic U-Net of \citep{deshpande_probabilistic_2022}  was trained to predict the displacement of a mesh given the force boundary values.
The predicted uncertainties of the network correlated well with the prediction errors.
\citet{dang_probabilistic_2021} used a Bayesian NN for structural reliability analysis.
They reported a reduction in computation time up to three orders of magnitude compared to a classical approach using Monte Carlo sampling with FEM. 
\citet{van_de_weg_long_2024} used a Bayesian approach to automatically determine the LSTM network size during training by capturing the uncertainties within the weights of the surrogate model. 
Their approach was immune to overfitting and showed accurate results as a surrogate model for solid mechanics. 
Other research used probabilistic NNs for material parameter identification \citep{rappel_tutorial_2020, zeraatpisheh_bayesian_2021} or for topology optimization \citep{patel_classification_2012}.

In this paper, we adapt the probabilistic NN architecture "Deep Variational Bayes Filter," introduced by \citet{karl_deep_2017-1} and enhanced by \citet{karl_unsupervised_2017}. 
DVBF extends Kalman Filter \citep{kalman_new_1960} to deep learning and improves the application to dynamical systems compared to \citet{krishnan_deep_2015} and \citet{johnson_composing_2017}.
The DVBF architecture used is mainly used in robotics \citep{becker-ehmck_learning_2020, karl_unsupervised_2017}.
While a DVBF-inspired architecture has been applied to fluid mechanics \citep{morton_parameter-conditioned_2021} and fluid control \citep{morton_deep_2018}, to the best of our knowledge, this is the first application of the DVBF to structural mechanics.

\section{Methodology}
\subsection{Finite Element Method Integrated Networks}

\begin{figure}
    \centering
    \begin{subfigure}[t]{\textwidth}
        \includegraphics{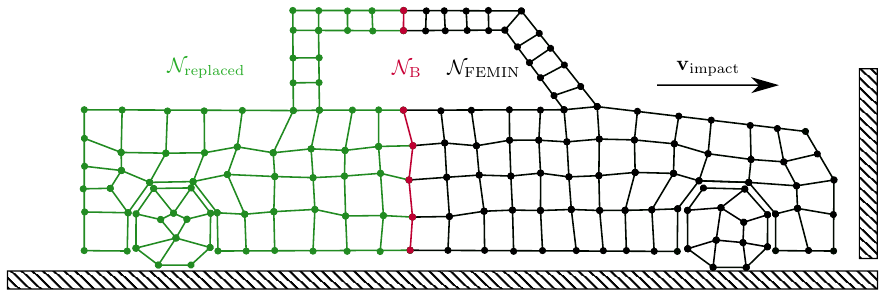}
        \caption{Mesh for data generation: Multiple FEM simulations are performed with design changes in $\mathcal{N}_{\text{FEMIN}}$. The displacement, velocity, and section forces are recorded at $\mathcal{N}_{\text{B}}$. }
    \end{subfigure}
    \begin{subfigure}[t]{\textwidth}
        \includegraphics{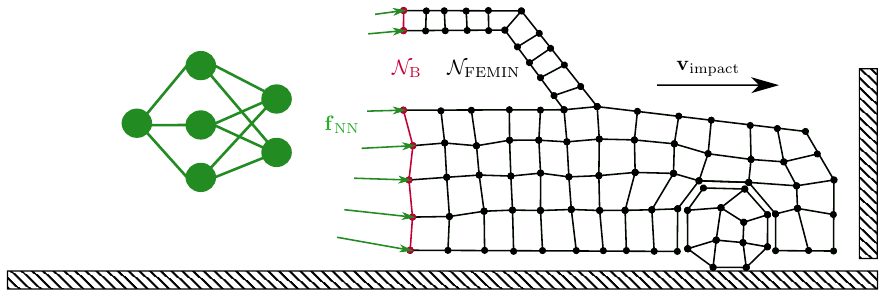}
        \caption{Mesh during FEMIN simulation: A NN replaces the nodes $\mathcal{N}_{\text{replaced}}$ by applying the force $\vec f_{\text{NN}}$ on $\mathcal{N}_{\text{B}}$.}
    \end{subfigure}
    \caption{Principle of FEMIN: A FEM simulation is accelerated by a NN, which replaces a portion of the FEM mesh.}
    \label{fig:FEMIN-principle}
\end{figure}

In FEMIN, a portion of the FEM domain is replaced by a NN, altering the original boundary value problem. 
The simplified equation of motion for highly dynamic FEM is \citep{pickett_fracture_1990}:
\begin{equation}
\tens{M}\vec{a} + \tens K \vec d = \vec f_{\text{ext}} \in \mathbb{R}^{n_{\text{FEM}} \cdot n_{\text{DOF}}}
\label{equ:orig-ODE}
\end{equation}
with\\[1.5ex]
\begin{tabular}{ll}
    $\tens{M}$ & mass matrix\\ 
    $\tens K$ & stiffness matrix\\
    $\vec d$ &  displacement vector\\
    $\vec a$ &  acceleration vector\\
    $ \vec f_{\text{ext}}$ & external force\\ 
    $n_{\text{FEM}}$ & total number of nodes in FEM simulation \\
    $n_{\text{DOF}}$ & number of degrees of freedom \\
\end{tabular}
\newline 
The FEM mesh is partitioned into two subsets: $\mathcal{N}_{\text{FEMIN}}$ (nodes retained during a FEMIN simulation) and $\mathcal{N}_{\text{replaced}}$ (nodes replaced by the NN). 
The boundary nodes $\mathcal{N}_{\text{B}}$ are a subset of $\mathcal{N}_{\text{FEMIN}}$. 
The modified boundary value problem for FEMIN becomes:
\begin{equation}
    \tens{M}_{\text{FEMIN}}\vec{a}_{\text{FEMIN}} + \tens K_{\text{FEMIN}} \vec d_{\text{FEMIN}} = \vec f_{\text{ext, FEMIN}} +  \vec f_{\text{NN}} \in \mathbb{R}^{n_{\text{FEMIN}} \cdot n_{\text{DOF}}}
    \label{equ:FEMIN-ODE}
\end{equation}
where $\vec f_{\text{NN}}$ is the force vector provided by the NN acting on the boundary nodes $\mathcal{N}_{\text{B}}$ and $n_{\text{FEMIN}}$ is the number of nodes in the FEMIN simulation
Since it is
\begin{equation}
  n_{\text{FEMIN}} < n_{\text{FEM}}\, ,  
\end{equation}
the boundary value problem for FEMIN can be solved with fewer computational resources. 
The NN must approximate the section forces $\vec{f}_{\text{B}}$ acting between $\mathcal{N}_{\text{FEMIN}}$ and $\mathcal{N}_{\text{replaced}}$:
\begin{equation}
     \vec{f}_{\text{NN}} \approx \vec{f}_{\text{B}}
\end{equation}

The prediction problem is characterized by:
\begin{itemize}
    \item Partial observability: The NN must infer $\vec{f}_{\text{B}}$ based solely on boundary kinematic ($\vec{d}_{\text{B}}$, $\vec{v}_{\text{B}}$).
    The node set $\mathcal{N}_{\text{replaced}}$ is not available during a FEMIN simulation and including more information from $\mathcal{N}_{\text{FEMIN}}$ would increase the complexity of the NN matrices, reducing and even negating the computational benefits of FEMIN.
    \item State dependency: The mechanical behavior depends on both spatial state ($\vec{d}$, $\vec{v}$) and system state ($\vec \gamma$). The system state $\vec \gamma$ describes properties such as plasticity, hardening, and damage. These variables are not necessarily nodal properties. Since the prediction problem follows a node-based description, the system state cannot improve the prediction.
\end{itemize}
Training data is obtained from full FEM simulations, extracting boundary node kinematics and loads:
\begin{equation}
    \vec{d}_{\text{B}, t}, \vec{v}_{\text{B}, t},  \vec{f}_{\text{B}, t} \in \mathbb{R}^{n_{\text{B}} \cdot n_{\text{DOF}}} \text{ with } t \in T\, 
\end{equation}
at every time step $t$. 
The total number of time steps is denoted as $T$ and the number of boundary nodes is $n_{\text{B}}$.
For data generation, design changes are made in $\mathcal{N}_{\text{FEMIN}}$. 
If the design changes stem from a parametric model, these parameters can be used to improve the prediction.
These parameters can be used as the parameter vector $\vec p$.
If the design changes are established manually or are not parametric (e.g., from Graph and Heuristic based topology optimization  \citep{ortmann_graph_2013} or a generative approach \citep{siddiqui_meshgpt_2023}), the parameter vector is  $\vec p = \vec 0$.

The prediction problem can now be written as a mapping between the kinematics to forces at time step $t$
\begin{equation}\label{equ:prediction-problem}
     f_{\text{NN}}\!: \quad \left(\vec{d}_{\text{B}, t}, \vec{v}_{\text{B}, t},   \vec{p} \right) \mapsto \vec f_{\text{NN}, t}\,, 
\end{equation}
where $f_{\text{NN}}$ described by an NN. 
The FEM solver handles the time step evolution via its integration scheme.,
The FEMIN principle is depicted by \cref{fig:FEMIN-principle}.

\subsection{Deep Variational Bayes Filter}
\subsubsection{Original Fusion DVBF}

The architecture is based on the Fusion DVBF presented by \citet{karl_unsupervised_2017}.
Displacement $\vec{d}_{t}$, velocity $\vec{v}_{t}$ and force $\vec{f}_{t}$ with time step $t$ are concatenated to observation $\vec{x}_{t}$:
\begin{equation}\label{equ:def-x}
    \vec{x}_{t} = \left\{ \vec{d}_{t}, \vec{v}_{t},  \vec{f}_{t} \right\} \in \mathbb{R}^{k}
\end{equation}
where $k$ is the dimensionality of the observations.
The encoder $q_\text{enc}$ encodes $\vec{x}_{t}$ to the latent state estimation $\vec{z}_{\text{enc}, t}$:
\begin{equation}
    q_\text{enc}(\vec{z}_{t, \text{enc}}\,\vert\,\vec{x}_{t} ) = \mathcal{N}(\vec{\mu}_{\text{enc}, t}(\vec{x}_{t}),\,\vec{\sigma}^2_{\text{enc}, t}(\vec{x}_{t})) \in \mathbb{R}^{l}
\end{equation} 
where $l$ is the dimensionality of the latent state. 
The mean $\vec{\mu}_{\text{enc}, t}(\vec{x}_{t})$ and the standard deviation $\vec{\sigma}_{\text{enc}, t}(\vec{x}_{t})$ of the Gaussian normal distribution are represented by a NN.
The transition model provides an estimate based on the previous latent state $\vec{z}_{t-1}$. 
The transition estimate of $\vec{z}_{t}$ can be enhanced by the design parameters $\vec{p}$ given their availability.
\begin{equation}
    p_{\text{trans}}(\vec{z}_{\text{trans}, t}\,\vert\,\vec{z}_{t-1}, \vec{p} ) = \mathcal{N}(\vec{\mu}_{\text{trans}, t}(\vec{z}_{t-1}, \vec{p}), \vec{\sigma}^2_{\text{trans}, t}(\vec{z}_{t-1}, \vec{p}))  \in \mathbb{R}^{l}\, .
\end{equation}
The prior is also given by the transition model.
\begin{equation}
        p_{\text{prior}}(\vec{z}_{\text{prior}, t}\,\vert\,\vec{z}_{t-1}, \vec{p} ) = \mathcal{N}(\vec{\mu}_{\text{trans}, t} (\vec{z}_{t-1}, \vec{p}), \vec{\sigma}^2_{\text{prior}, t} (\vec{z}_{t-1}, \vec{p}))  \in \mathbb{R}^{l} \, .
\end{equation}
While the mean $\vec{\mu}_{\text{trans}}$ is shared between the transition estimate of $\vec{z}_t$ and the prior, the standard deviations of the two distributions $ p_{\text{trans}}$ and $ p_{\text{prior}}$ distribution are different.
Using two different standard deviations for prior and transition improves optimization of the KL-divergence \citep{karl_unsupervised_2020}.

The transition model is based on the locally linear transition from \citet{karl_deep_2017-1}.
\citet{watter_embed_2015} demonstrated the superior performance of the locally linear transition for long-term prediction problems while maintaining computational efficiency. 
The mean of the transition estimate and the prior is given by 
\begin{equation} \label{equ:locally-linear-transition}
        \vec{\mu}_{\text{trans}, t} = \vec{z}_{t-1} + \tens{A}_{t} \vec{z}_{t-1} + \tens{B}_{t} \vec{p}
\end{equation}
The matrices $\tens{A}_{t}$ and $\tens{B}_{t}$ are constructed from base matrices:
\begin{align}
    \tens{A}_{t} =\sum\limits_{i=1}^{M} \alpha_{t}^{(i)}  \tanh_{s}\left( \tens{A}^{(i)}  \right)\,, \qquad &
    \tens{B}_{t} = \sum\limits_{i=1}^{M} \alpha_{t}^{(i)}  \tanh_{s}\left( \tens{B}^{(i)} \right)
\end{align}
where $M$ denotes the number of linear systems.
The components of the base matrices  $\tens{A}^{(i)}$ and $\tens{B}^{(i)}$ are learned during training. 
The mixing factor $\vec \alpha_{t}$ is obtained from a neural network $f_{\alpha}$ with final sigmoid activation:
\begin{equation}\label{equ:alpha-network}
    \vec{\alpha}_{t} = f_{\alpha}(\vec{z}_{t-1}, \vec{p}) \in \mathbb{R}^{M} \, .
\end{equation}
The scaled hyperbolic tangent limits the maximum values of the base matrices
\begin{equation}
    \tanh_{s}(x) = s \tanh x
\end{equation}
with scale $s$. 
Note that, while the vector $\vec p$ is constant over time, the matrix-vector product $\tens{B}_{t} \vec{p}$ in \cref{equ:locally-linear-transition} is not because mixing factor $\vec \alpha_t$ depends on the latent state $\vec z _{t-1}$ (c.f. \cref{equ:alpha-network}).
Thus, the inclusion of the matrix $\tens B_t$ into the transition model controls when the information from the design parameters is needed. 
The standard deviations of the transition model are
\begin{align}
    \vec{\sigma}_{t, \text{trans}} = \sum\limits_{i=1}^{M} \alpha_{t}^{(i)} \psi \left( \vec{c}_{1}^{(i)}\right) \, , \qquad &
    \vec{\sigma}_{t, \text{prior}} = \sum\limits_{i=1}^{M} \alpha_{t}^{(i)} \psi \left( \vec{c}_{2}^{(i)}\right)
\end{align}
with the base vectors $\vec{c}_{1}^{(i)}$ and $\vec{c}_{2}^{(i)}$ and the softplus function $\psi$ ensuring positive values
\begin{equation}
    \psi(x) = \log \left( e^x +1 \right) \,.
\end{equation}

The posterior distribution combines the transition $p_{\text{trans}}$ and encoder distributions $q_{\text{enc}}$. It follows the Markov assumption
\begin{equation}
    q(\vec{z}_{t}\vert \vec{z}_{1:t-1}, \vec{x}_{1:t}, \vec{p}) \propto   q_\text{enc}(\vec{z}_{\text{enc}, t} \vert \vec{x}_{t} ) \times p_{\text{trans}}(\vec{z}_{\text{trans}, t} \vert  \vec{z}_{t-1}, \vec{p} ) = \mathcal{N}(\vec{\mu}_{q, t}, \vec{\sigma}^2_{q, t}) 
\end{equation}
with
\begin{align}
    \vec{\mu}_{q, t} =  \frac{\vec{\mu}_{\text{enc}, t}  \vec{{\sigma}}^2_{\text{trans}, t} + \vec{\mu}_{\text{trans}, t}  \vec{\sigma}^2_{\text{enc}, t} }{\vec{\sigma}^2_{\text{enc}, t} + \vec{{\sigma}}^2_{\text{trans}, t}}, \qquad & 
    \vec{\sigma}^2_{q, t} = \frac{\vec{\sigma}^2_{\text{enc}, t} \vec{{\sigma}}^2_{\text{trans}, t} }{\vec{\sigma}^2_{\text{enc}, t} + \vec{{\sigma}}^2_{\text{trans}, t}} \, .
\end{align}
The latent state is sampled from this posterior distribution
\begin{equation}\label{equ:post-sampling}
    \vec{z}_t \sim \mathcal{N}(\vec{\mu}_{q, t}, \vec{\sigma}^2_{q, t}) \, .
\end{equation}
The reparametrization trick \citep{kingma_auto-encoding_2014} allows backpropagation through this sampling process
The decoder provides the likelihood distribution
\begin{equation}\label{equ:decoding}
 p_{\text{lik}}(\vec{\hat{x}}_{t} \vert \vec{z}_t) = \mathcal{N}(\vec{\mu}_{\text{lik}, t}, \vec{\sigma}^2_{\text{lik}, t})\, .
\end{equation}
The DVBF is trained by minimizing the negative Evidence Lower Bound (ELBO) function
\begin{equation}
    \mathcal{L}_{\text{model}} = - \frac{1}{N}\sum\limits_{n=1}^{N}\mathcal{L}_{\text{ELBO}}\left(\vec{x}_{1:T}^{(n)}, \vec{p}^{(n)}, \vec{\lambda}\right) := - \mathcal{L}_{\text{recon}} + \mathcal{L}_{\text{KL}}
\end{equation}
with $N$ number of samples.
Backpropagation through time provides the gradients necessary to perform stochastic gradient decent optimizing $\mathcal{L}$ by varying the trainable parameters $\vec{\lambda}$. 
The ELBO loss consists of the negative log-likelihood as the reconstruction loss
\begin{equation}\label{equ:nll}
   - \mathcal{L}_{\text{recon}, t} = - \frac{1}{2} \sum_{i=1}^k \left(  \ln(\sigma_{\text{lik}, t, i}^2) - \frac{(x_{t, i} - \mu_{\text{lik}, t, i})^2}{\sigma_{\text{lik}, t, i}^2} \right) -\frac{k}{2}\ln(2\pi)
\end{equation}
and Kullback Leibler (KL) divergence
\begin{equation}\label{equ:kl}
    \mathcal{L}_{\text{KL}, t} = D_{\text{KL}}\left(q(\vec{z}_t\vert \vec{x}_t, \vec{p}) \parallel p_{\text{prior}}(\vec{z}_{t} \vert \vec{p})\right) = \frac{1}{2} \sum_{i=1}^l \left(\frac{\sigma_{q, t, i}^2}{\sigma_{\text{prior}, t, i}^2} + \frac{(\mu_{q, t, i} - \mu_{\text{trans}, t, i})^2}{\sigma_{\text{prior}, t, i}^2}  + \ln \left(\frac{\sigma_{\text{prior}, t, i}^2}{\sigma_{q, t, i}^2}\right) - 1 \right) \, . 
\end{equation}
The initial prior is a standard normal distribution
\begin{equation}
    p_{\text{prior}, 1}= \mathcal{N}(\vec 0,\vec 1) \in \mathbb{R}^l \,.
\end{equation}
The loss terms $ \mathcal{L}_{\text{recon}, t}$ and $\mathcal{L}_{\text{KL}, t}$ are aggregated over time via average. 

\subsubsection{Modification of Fusion DVBF for FEMIN}

\begin{figure}
    \centering
    \includegraphics{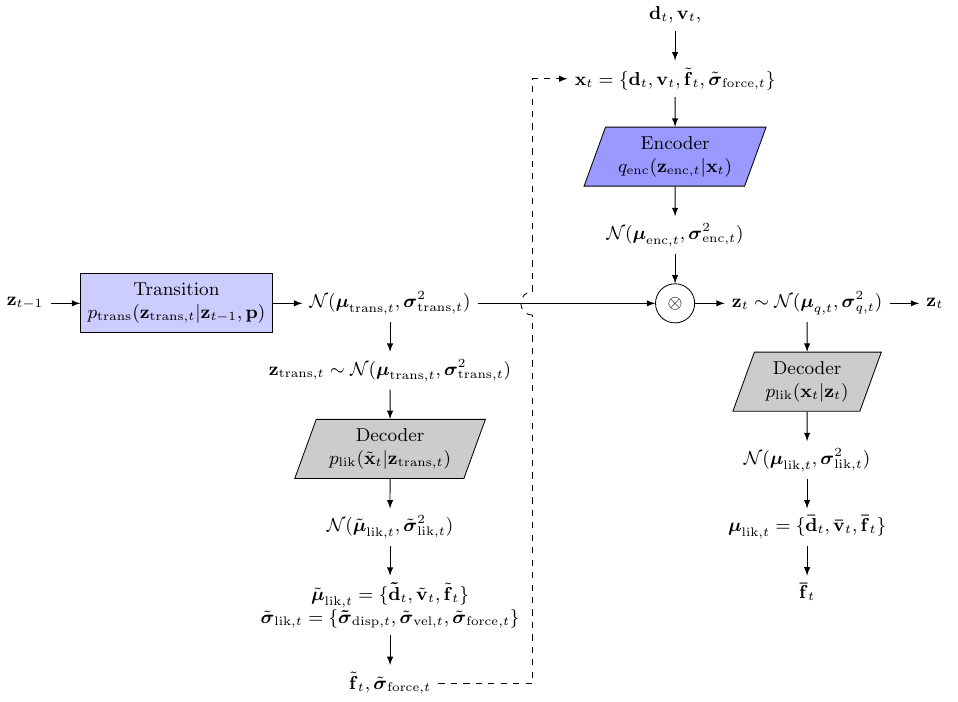}
    \caption{Flow chart of the modified DVBF for FEMIN. No backpropagation over dashed lines}
    \label{fig:dvbf}
\end{figure}

DVBF is a generative model, meaning its input and output are on the same manifold. 
This generative approach allows for holistic learning, using all available information to build the latent space.
However, our problem involves an FE solver that enforces a mapping between two distinct manifolds (kinematic to forces, see \cref{equ:prediction-problem}). 
To address this, we propose modifying the DVBF architecture to maintain holistic learning while respecting the mapping relationship imposed by the solver.

Our approach involves decoding the transition estimate of $\vec{z}_t$ provided by the transition model. Initially, a latent state estimate is sampled from the transition distribution:
\begin{equation}\label{equ:pre-sampling}
    \vec{\Tilde{z}}_{\text{trans}, t} \sim  \mathcal{N}(\vec{\mu}_{\text{trans}, t}, \vec{\sigma}^2_{\text{trans}, t})
\end{equation}
This sample then undergoes a preliminary decoding step:
\begin{equation}
    p_{\text{lik}}(\vec{\Tilde{x}}_{t} \vert \vec{\Tilde{z}}_t) = \mathcal{N}(\vec{\Tilde{\mu}}_{\text{lik}, t}, \vec{\Tilde{\sigma}}^2_{\text{lik}, t})
\end{equation}
From the resulting distribution, we extract the components of $\vec{\Tilde{\mu}}^2_{\text{lik}}$ and $\vec{\Tilde{\sigma}}^2_{\text{lik}}$ that correspond to the force prediction. 
These extracted values replace the true force values in the input definition of \cref{equ:def-x}, redefining the DVBF input as:
\begin{equation}\label{equ:redef-x}
    \vec{x}_{t} = \left\{ \vec{d}_{t}, \vec{v}_{t}, \vec{\Tilde{\mu}}_{\text{lik, force}, t}, \vec{\Tilde{\sigma}}_{\text{lik, force}, t}\right\} \,.
\end{equation}

This modification eliminates the need for true force as input to the model while allowing the DVBF to learn from all three quantities: displacement, velocity, and force. 
When the DVBF is coupled with the FEM solver (also referred to as "online application"), the preliminary force prediction closely approximates the actual force of the final decoding step applied to the solver. 
Consequently, the input to the encoder — combined with the kinematics updated by the FEM solver — closely resembles the actual observation of the boundary, enhancing the model's accuracy.
The modified DVBF architecture is depicted in \cref{fig:dvbf}.

The use of both mean and standard deviation from the decoder to describe the force distribution provides valuable information to the model. 
This approach informs the encoder about the model's certainty in its force predictions, enabling it to weigh its belief more heavily on either the kinematic or the predicted force. 
Another approach to implicitly provide the encoder with information about the model's certainty in its force predictions would be to sample from preliminary force distribution. 
However, during the initial training phases, the likelihood model's standard deviations are typically large, reflecting the model's lack of confidence in its outputs. 
Sampling from such a broad distribution introduces excessive noise, potentially over-regularizing the model and reducing convergence. 
Therefore, our approach of using both the mean and standard deviation strikes a balance between leveraging the full force distribution information and maintaining good learning performance.

Gradients are not propagated through the preliminary decoding step and are stopped at \cref{equ:redef-x}.
This design choice offers two advantages.
Firstly, it improves training speed and memory efficiency by reducing the number of backpropagation steps.
Secondly, it enhances online accuracy. 
By stopping the gradients, we effectively treat $\vec{\Tilde{\mu}}_{\text{lik, force}}$ and $\vec{\Tilde{\sigma}}_{\text{lik, force}}$ as standard inputs and not trainable parameters.
The decoder weights are only updated with respect to the final decoding step (c.f. \cref{equ:decoding}).
Since the output of this decoding step is applied to the FEM solver during the online application, placing greater emphasis on the final decoding step improves online accuracy. 

\subsubsection{Window-based training}
Following the approach of \citet{thel_introducing_2024}, we implement window-based training. 
This method is particularly useful for FEMIN, where time series can be extremely long. 
Window-based training allows for effective backpropagation on smaller time series segments, mitigating issues with vanishing or exploding gradients \citep{bengio_learning_1994}. 
Additionally, this approach will scale well to even larger load cases.

The large time series with $T$ time steps is sliced into smaller, rolling windows of length $T^*$. 
An initial network is applied to the first $K$ time steps in each window:
\begin{equation}
    q_{\text{init}}(\vec{w}_0\vert \vec{x}_{1:K}, \vec{p}) = \mathcal{N}(\vec{\mu}_{w}, \vec{\sigma}^2_{w})
\end{equation}
Note that $q_{\text{init}}$ still receives the true force values $\vec{f}_{1:K}$.
We then sample from this initial distribution
\begin{equation}\label{equ:init-sampling}
    \vec{w}_0 \sim \mathcal{N}(\vec{\mu}_{w}, \vec{\sigma}^2_{w})
\end{equation}
and apply an initial nonlinear transition
\begin{equation}
    \vec{z}_1 = f_{\text{trans}, 0}(\vec{w}_0) \, , 
\end{equation}
where $f_{\text{trans}, 0}$ is a NN \citep{karl_deep_2017-1}.

\subsubsection{Online inference}

During the initialization of the online application, the input of $q_{\text{init}}$ are the initial values of the simulation padded $K$-times.
Since these are non-physical values, it can be expected that the predicted force during the initial time steps deviates from the ground truth values and converge during the simulation's progress.

The sampling processes in \cref{equ:post-sampling,equ:pre-sampling,equ:init-sampling} are replaced by directly choosing the mean of these distributions.
While this does not mean that these lead to the most probable outcomes in the likelihood distribution of \cref{equ:decoding} (the mapping from posterior to likelihood is nonlinear), it leads to a smooth latent trajectory and repeatable results.
The smooth latent state also leads to a smooth force trajectory, reducing vibrations during online application. 
Similarly, the force that is applied to the FEM solver is the mean of the force distribution of the likelihood model $p_{\text{lik}}(\vec{x}_t|\vec{z}_t)$.

\section{Application}
\subsection{Load cases}

\begin{figure}
    \centering
    \begin{subfigure}[t]{0.4\textwidth}
        \includegraphics[scale=0.5]{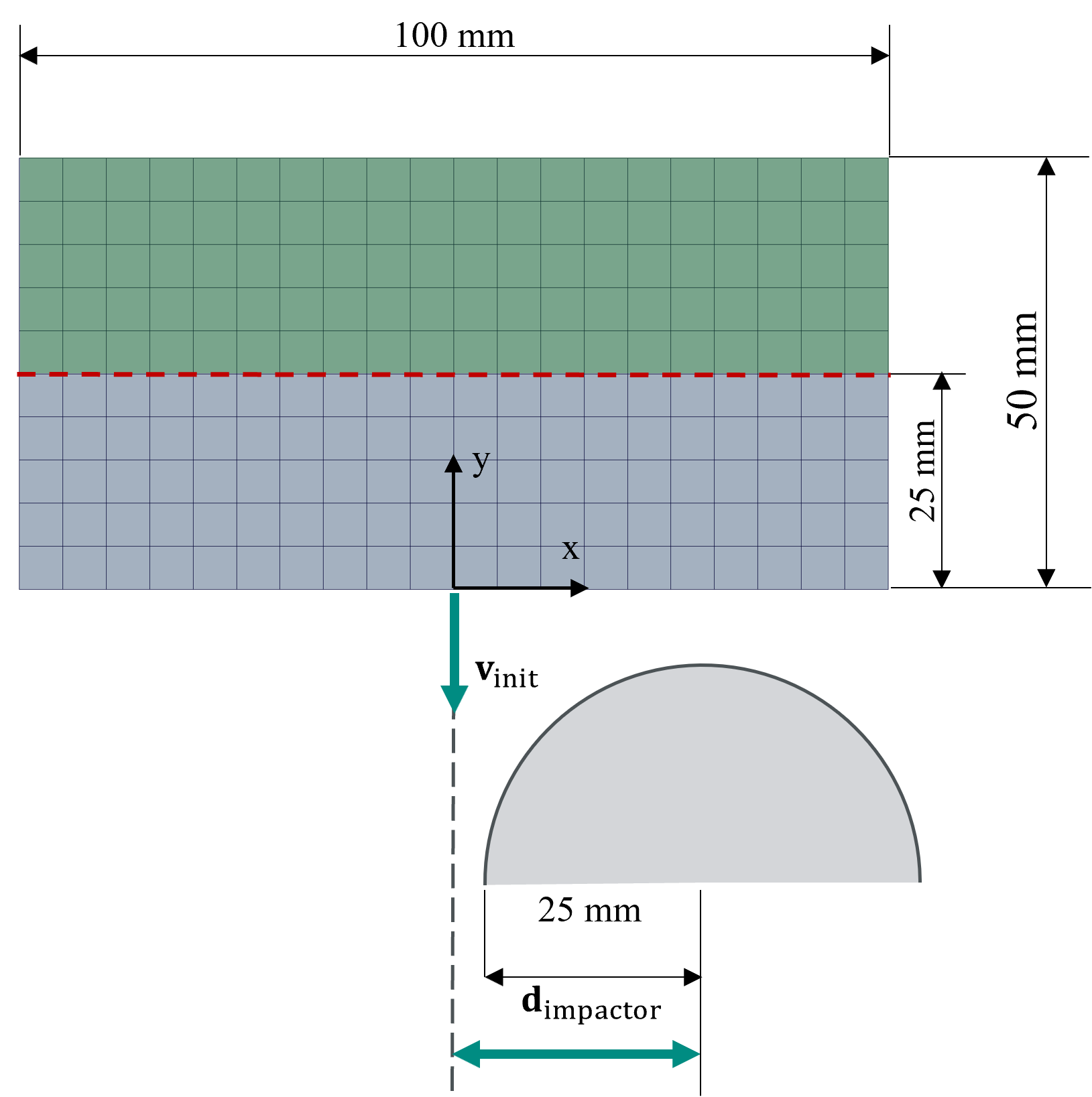}
        \caption{Box-Impact (BI) load case}
    \end{subfigure}
    \begin{subfigure}[t]{0.5\textwidth}
        \includegraphics[scale=0.5]{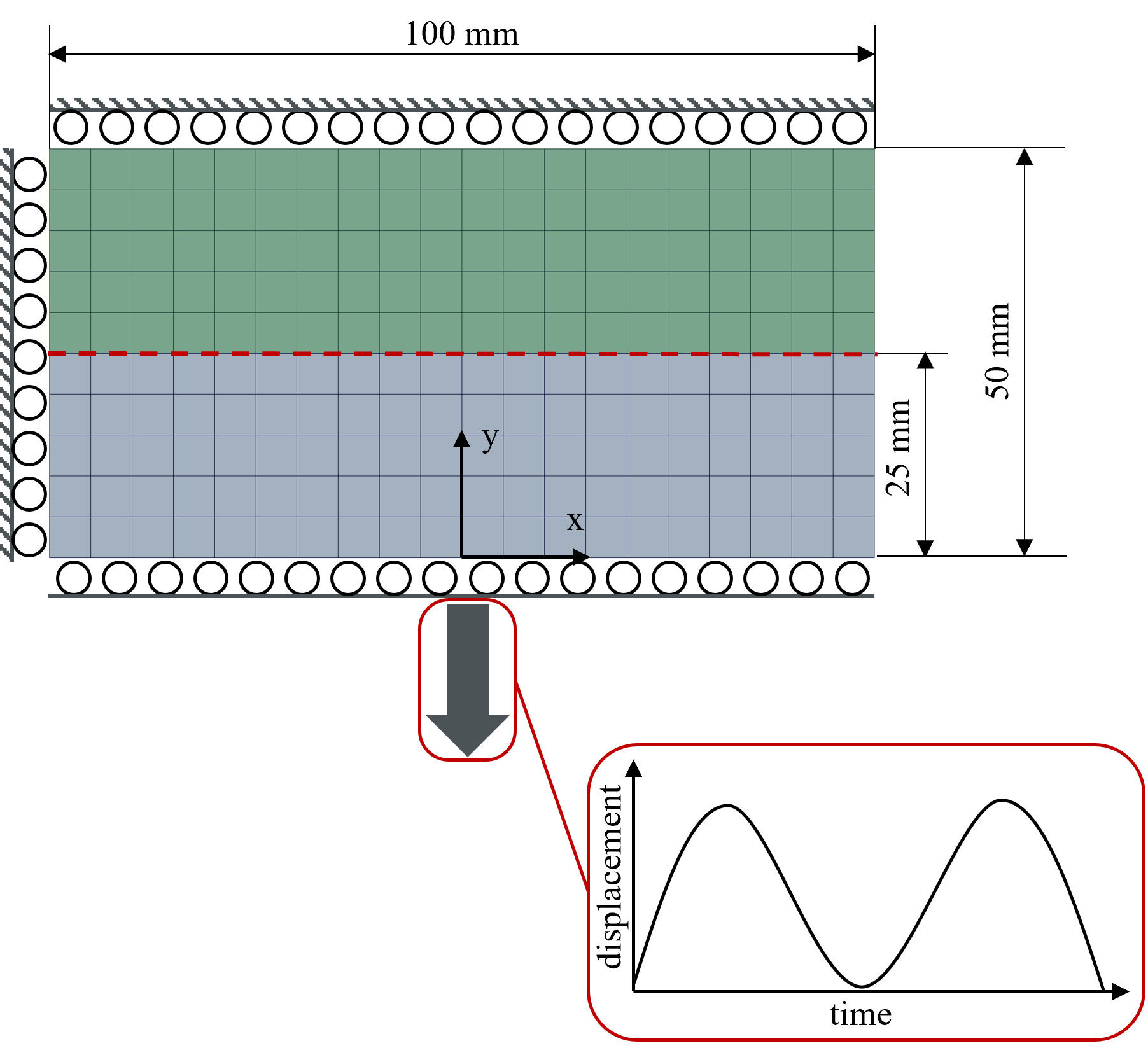}
        \caption{Tension-Compression-Tension (TCT) load case}
    \end{subfigure}
    \caption{Load cases as presented by \citep{thel_introducing_2024}. The elements depicted in green are in the node set $\mathcal{N}_{\text{replaced}}$, and the gray elements are in $\mathcal{N}_{\text{FEMIN}}$. The boundary nodes $\mathcal{N}_{\text{B}}$ are marked with a red dashed line. }
    \label{fig:load-cases}
\end{figure}

This paper utilizes the same load cases as \citet{thel_introducing_2024}, which we briefly summarize here for convenience.  
Two distinct load cases -- "Box Impact" (BI) and "Tension-Compression-Tension" (TCT) --  were simulated using a parameterized finite element model (see \cref{fig:load-cases}).
Both cases employed a \SI{100}{\mm} x \SI{50}{\mm} x \SI{3.5}{\mm} sheet metal specimen (22MnB5 steel, \citep{eller_modeling_2016}) discretized with shell elements with element edge length of \SI{5}{\mm}.
For each load case, 22 designs were generated using the explicit FEM solver VPS \citep{esi_group_virtual_2021}, with 16 for training, 3 for validation, and 3 for testing. 
Simulations ran for 3 ms with a time step of \SI{0.5e-3}{\milli\s} totalling $T=6000$ time steps. 
The displacement, velocity, and section force were extracted from the $n_{\text{B}} = 21$ interface nodes along the specimen's midline.

The BI load case is an academic representation of a typical crash scenario with the specimen impacting a rigid half-circular structure. 
The design parameters are the specimen's initial velocity in y-direction (ranging from \SI{-30}{\m\per\s} to \SI{-10}{\m\per\s}) and the impactor's x-position (ranging from \SI{-15}{\mm} to \SI{15}{\mm}).

The TCT load case aims to capture history-dependent material behavior by applying a sinusoidal displacement-controlled load at the specimen's lower edge. 
The maximum displacement is large enough so that the specimen reaches plasticity.
In the next load cycle, the specimen undergoes the same deformation pattern as in the previous cycle, but the force levels change due to plastic strain hardening. 
The loading frequency (ranging from \SI{500}{\hertz} to \SI{2000}{\hertz}) is not included in the parameter vector, so it is $\vec p = \vec 0$.

\subsection{Metrics}

The data set split into the train, validation, and test data is the same as \citep{thel_introducing_2024}, enabling direct comparison of results. 
We evaluate accuracy and confidence metrics, reporting them on the test data set. 
These metrics are computed between the predicted quantity of interest (QoI) $\hat{\vec{\chi}}_{n,t} \in \mathbb{R}^C$ and its true value $\vec{\chi}_{n,t}$ with the corresponding predicted confidence denoted as $\vec{\sigma}_{n,t} \in \mathbb{R}^C$

For accuracy assessment, we use two primary metrics. 
First, the mean squared error (MSE):
\begin{equation}
    \text{MSE} =  \frac{1}{N\cdot T \cdot C}\sum\limits_{n=1}^{N} \sum\limits_{t=1}^{T} \sum\limits_{c=1}^{C} ( \chi_{n,t,c} - \hat{\chi}_{n,t,c})^2 \,.
\end{equation}
Second, the coefficient of determination, or $R^2$ score, is defined as:
\begin{equation}
R^2_{n,t} = 1 - \frac{\sum_{c=1}^C (\chi_{n,t,c} - \hat{\chi}_{n,t,c})^2}{\sum_{c=1}^C (\chi_{n,t,c} - \bar{\chi}_{n,t})^2}
\end{equation}
where $\bar{\chi}_{n,t}$ is the mean of the true values per time step $t$ and sample $n$. 
The $R^2$ score ranges from $-\infty$ to $1$, with $1$ indicating perfect prediction. 
This normalized metric allows for comparison across different load cases. 
To account for variations in prediction accuracy throughout the simulation, we employ the variance-weighted $R^2$ score:
\begin{equation}
R^2= \sum_{n=1}^N \sum_{t=1}^T \left(\frac{\text{var}(\vec \chi_{n,t})}{\sum_{n=1}^N \sum_{t=1}^T \text{var}(\vec \chi_{n,t})} R^2_{n,t}\right)
\end{equation}
This approach reduces the impact of small prediction errors in low-variance regions (e.g., during the beginning of the simulation), which could otherwise disproportionately affect the overall score.

For confidence assessment, we employ three metrics. 
The negative log-likelihood (NLL) is the reconstruction loss of the DVBF 
\begin{equation}
    \text{NLL} = - \mathcal{L}_{\text{recon}}
\end{equation}
The NLL weights prediction accuracy with prediction confidence. 
A prediction that is inaccurate while the model is confident (low predicted uncertainty) will result in a larger value than an inaccurate prediction with low confidence. 
Therefore, smaller values are desirable. 
Note that the NLL is not bounded in either direction and can become negative.
Another confidence metric is the Prediction Interval Coverage Probability (PICP), which offers a human-interpretable confidence score:
\begin{equation}
   \text{PICP} = \frac{1}{N\cdot T \cdot C}\sum\limits_{n=1}^{N} \sum\limits_{t=1}^{T} \sum\limits_{c=1}^{C}  \xi_{i,t,c}, \qquad     \xi_{i,t,c} = \begin{cases}
      1, & \text{if}\ \chi_{i,t,c} \in [\hat{\chi}_{i,t,c} - 1.96 \sigma_{i,t,c}, \quad \hat{\chi}_{i,t,c} + 1.96 \sigma_{i,t,c}] \\
      0, & \text{otherwise}
    \end{cases}  
\end{equation}
The PICP is the proportion of true values that fall within the \SI{95}{\percent}-confidence interval, with higher values indicating better confidence estimation.
To evaluate the model's overall confidence, we calculate the average width of the confidence band:
\begin{equation}
    \text{AC} = \frac{2 \cdot 1.96}{N\cdot T \cdot C}\sum\limits_{n=1}^{N} \sum\limits_{t=1}^{T} \sum\limits_{c=1}^{C}  \sigma_{i,t,c} \, .
\end{equation}
The average confidence (AC) value is reported in the same unit as the quantity of interest $\hat{\vec{y}}$.

Finally, to assess the correlation between two curves $\vec \phi \in \mathrm{R}^n$ and $\vec \eta \in \mathrm{R}^n$, we use the Pearson correlation coefficient $r$:
\begin{equation}
    r =\frac{\sum^n_{i=1}(\phi_i - \bar{\phi})(\eta_i - \bar{\eta})}{\sqrt{\sum ^n_{i=1}(\phi_i - \bar{\phi})^2} \sqrt{\sum ^n_{i=1}(\eta_i - \bar{\eta})^2}} \, ,
\end{equation}
where $\bar{\phi}$ and $\bar{\eta}$ are the respective means. 
A Pearson correlation of 1 indicates a perfect linear correlation.

\subsection{Results}

The hyperparameters used to generate these results are given in \cref{sec:hyperparameter}.

\subsubsection{Box Impact (BI)}

\begin{figure}
    \centering
    \begin{subfigure}[t]{0.33\textwidth}
        \includegraphics[width=\textwidth]{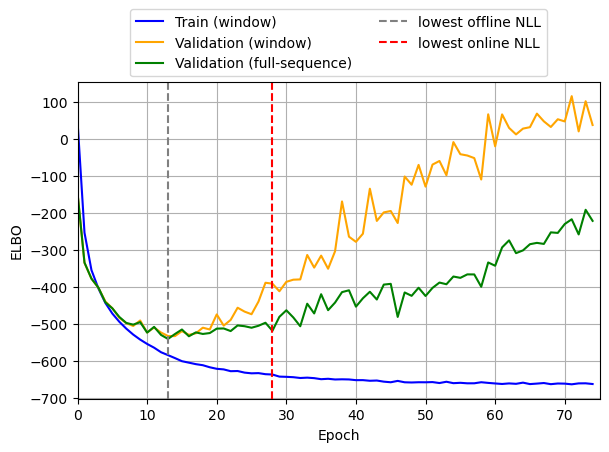}
        \label{fig:bi_elbo}
        \caption{ELBO (offline)}
    \end{subfigure}
    \hfill
    \begin{subfigure}[t]{0.33\textwidth}
        \includegraphics[width=\textwidth]{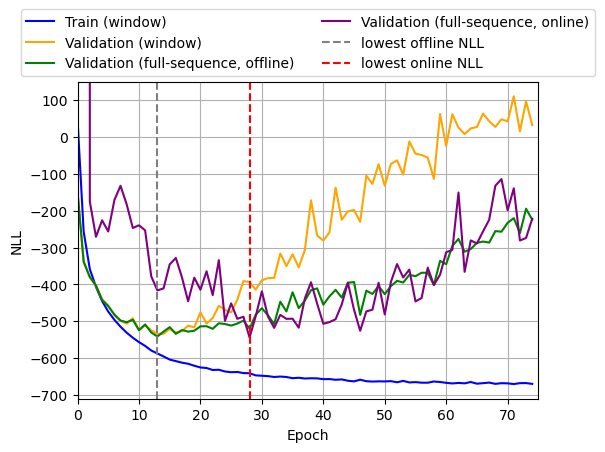}
        \label{fig:bi_nll}
        \caption{NLL (offline / online)}
    \end{subfigure}
    \hfill
    \begin{subfigure}[t]{0.33\textwidth}
        \includegraphics[width=\textwidth]{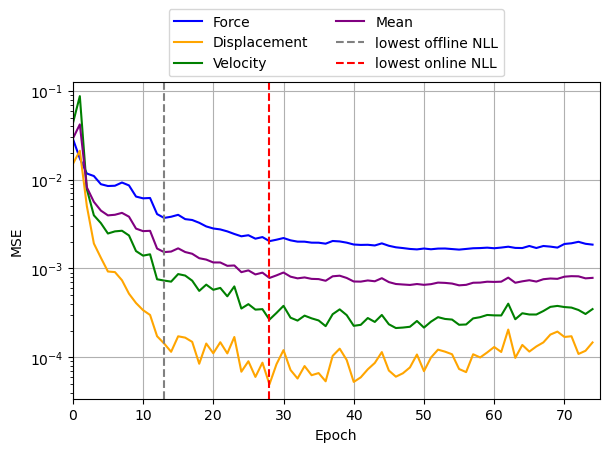}
        \label{fig:bi_mse}
        \caption{MSE (online)}
    \end{subfigure}
    \caption{Comparison of ELBO, NLL, and MSE over epochs for BI load case with $\vec p = \{\vec{v}_{\text{init}}, \vec{d}_{\text{impactor}}\}$}
    \label{fig:bi-loss}
\end{figure}

\begin{figure}
    \centering
    \begin{subfigure}[t]{0.33\textwidth}
        \includegraphics[width=\textwidth]{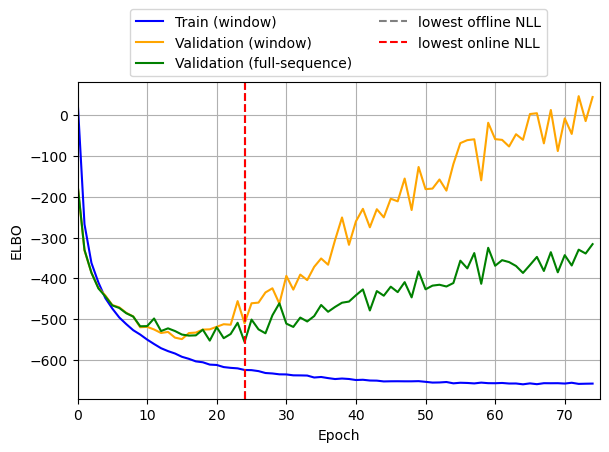}
        \label{fig:bi_np_elbo}
        \caption{ELBO (offline)}
    \end{subfigure}
    \hfill
    \begin{subfigure}[t]{0.33\textwidth}
        \includegraphics[width=\textwidth]{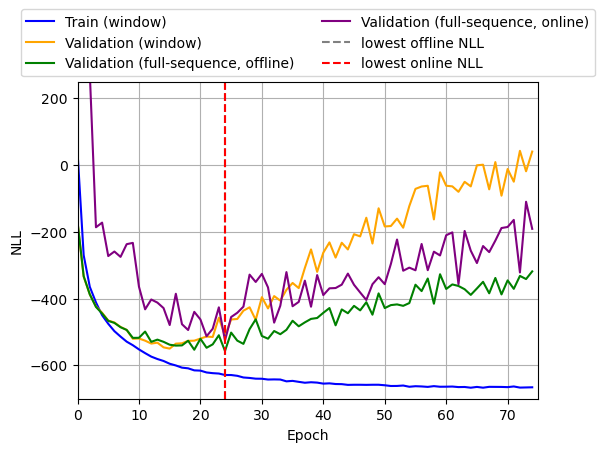}
        \label{fig:bi_np_nll}
        \caption{NLL (offline / online)}
    \end{subfigure}
    \hfill
    \begin{subfigure}[t]{0.33\textwidth}
        \includegraphics[width=\textwidth]{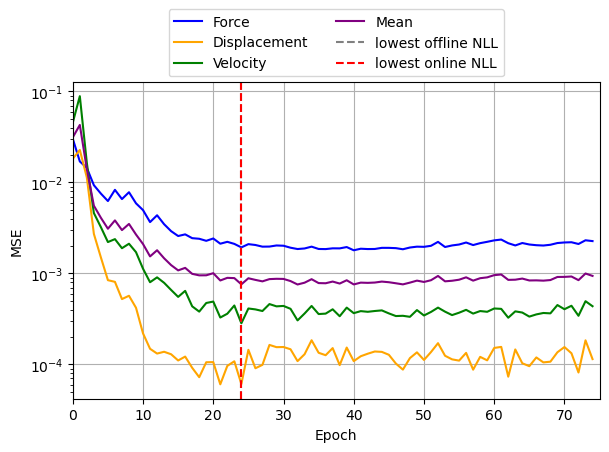}
        \label{fig:bi_np_mse}
        \caption{MSE (online, validation)}
    \end{subfigure}
    \caption{Comparison of ELBO, NLL, and MSE over epochs for BI load case ($\vec p = \vec 0$)}
    \label{fig:bi-loss-np}
\end{figure}

The training characteristics for both configurations ($\vec p = \{\vec{v}_{\text{init}}, \vec{d}_{\text{impactor}}\}$ and $\vec p = \vec 0$) exhibit similar patterns, as illustrated in \cref{fig:bi-loss,fig:bi-loss-np}. 
The offline metrics indicate that the models start overfitting after epoch 15; after that, the validation offline ELBO/NLL increases. 
Contrary to that, the online MSE decreases throughout the training, suggesting that the model overfits by becoming overconfident rather than less accurate. 
The DVBF demonstrates excellent generalization from window-based training to full-sequence evaluation, as the NLL evaluated on a window with length $T^*=300$ shows the same minimum and training evolution as the NLL evaluated on the full-sequence.
This illustrates the effectiveness of the window-based training. 

The weights for the subsequent analysis were chosen based on the lowest online NLL for the validation set. 
This approach is concurrent with \citet{thel_introducing_2024}, where the weights were chosen based on the lowest online MSE.
The epochs with the lowest online NLL occurred at epoch 28 for $\vec p = \{\vec{v}_{\text{init}}, \vec{d}_{\text{impactor}}\}$ and at epoch 24 for  $\vec p = \vec 0$.
It should be noted that choosing the weights based on an online metric might not be feasible for larger load cases where even the accelerated FEMIN simulation could take several hours.
If the weights were to be chosen based on the lowest offline NLL, the online performance would be slightly worse. 
In the case of the configuration $\vec p = \{\vec{v}_{\text{init}}, \vec{d}_{\text{impactor}}\}$, the online NLL would then be -416.08 in epoch 13 instead of -544.0 in epoch 28.
The configuration with $\vec p  = \vec 0$ does not show this effect.
Here, the epoch with the lowest offline NLL is also the epoch with the lowest online NLL.
These results show that choosing the DVBF's weights based on an offline metric does not severely -- if at all -- impact the online performance. 
A similar effect was observed by \citet{thel_introducing_2024} for the training of the LSTM networks. 

\begin{figure}
    \centering
    \includegraphics[width=\textwidth]{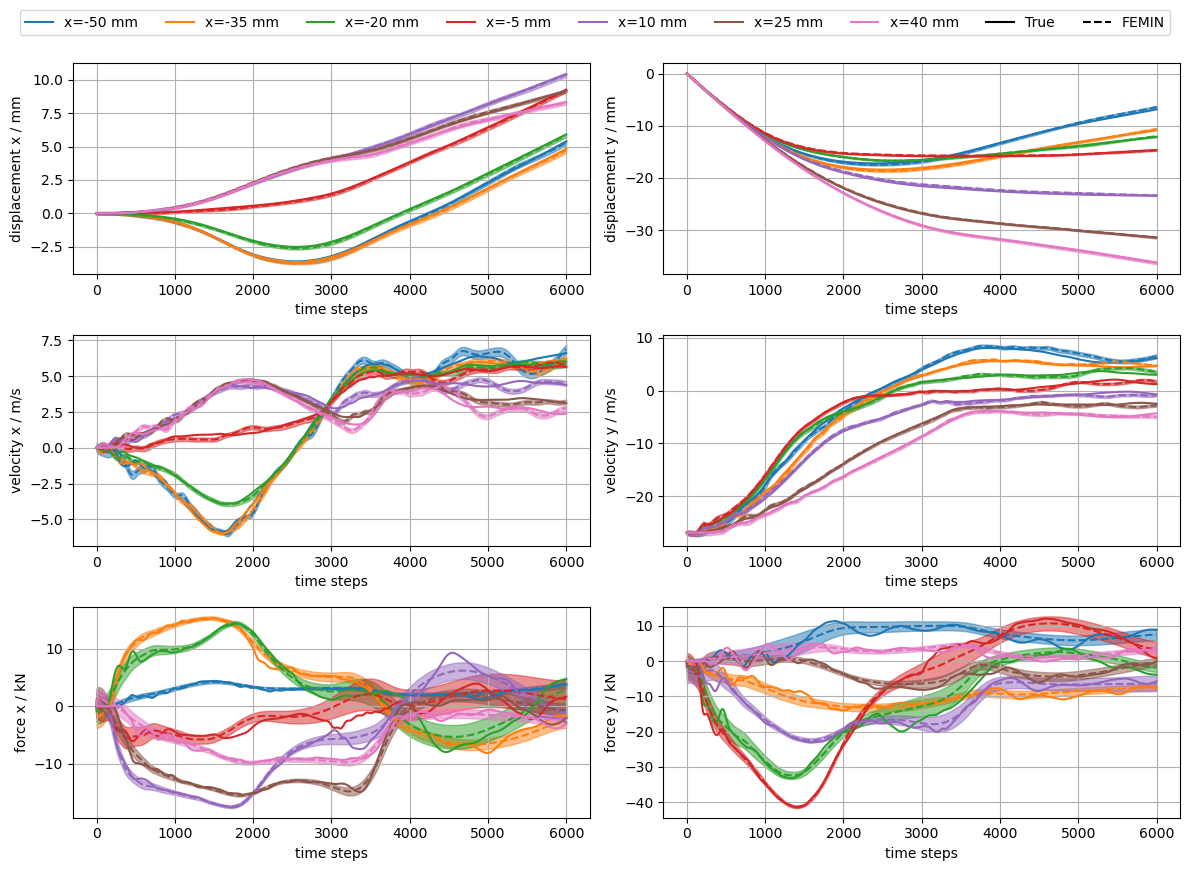}
    \caption{Online/FEMIN results of BI load case with $\vec p = \{\vec{v}_{\text{init}}, \vec{d}_{\text{impactor}}\}$ for one test sample. The FEMIN run (dashed lines) is compared with the full FEM simulation (continuous lines). The results are given for selected nodes at the boundary $\mathcal{N}_{\text{B}}$. The shaded areas represent the standard deviation computed by the decoder.}
    \label{fig:bi-femin-curves}
\end{figure}

\begin{figure}
    \centering
    \includegraphics[width=\textwidth]{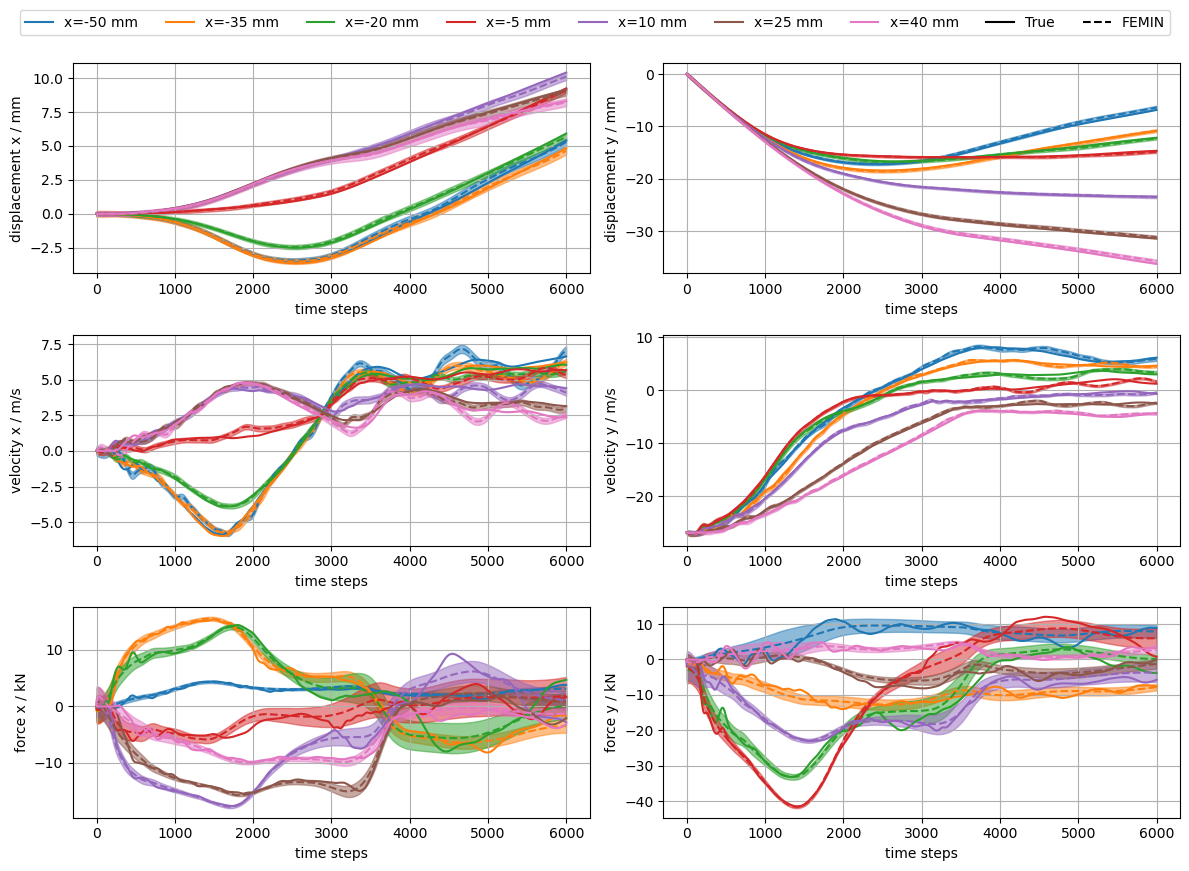}
    \caption{Online/FEMIN results of BI load case with $\vec p = \vec 0$ for one test sample. The FEMIN run (dashed lines) is compared with the full FEM simulation (continuous lines). The results are given for selected nodes at the boundary $\mathcal{N}_{\text{B}}$. The shaded areas represent the standard deviation computed by the decoder. } 
    \label{fig:bi-np-femin-curves}
\end{figure}

\begin{table}
\caption{Accuracy metrics of FEMIN online runs compared to the test set of the BI load case with $\vec p = \{\vec{v}_{\text{init}}, \vec{d}_{\text{impactor}}\}$. Values are presented as MSE / $R^2$ score. Results of the LSTM networks are from \citet{thel_introducing_2024}.}
\label{tab:bi-test-performance}
\centering
\begin{tabular}{@{}lccc@{}}
\toprule
\multicolumn{1}{c}{\textbf{QoI}} &\multicolumn{1}{c}{\textbf{DVBF}} & \multicolumn{1}{c}{\textbf{LSTM full-seq.}}      & \multicolumn{1}{c}{\textbf{LSTM window}} \\ \midrule
\textbf{Displacement}         & \textbf{\num{ 9.4171e-05}}  / \textbf{\num[round-mode=places,round-precision=3]{0.9976761407162412}}   & \num{1.3862e-04} / \num[round-mode=places,round-precision=3]{0.9965792831851102} & \num{3.1077e-04}  / \num[round-mode=places,round-precision=3]{0.9923310219663768}  \\
\textbf{Velocity}             & \num{4.2695e-04} / \textbf{\num[round-mode=places,round-precision=3]{0.991653160664581}} & \textbf{\num{4.1676e-04}} / \textbf{\num[round-mode=places,round-precision=3]{ 0.9918524381755355}}  & \num{8.2863e-04} / \num[round-mode=places,round-precision=3]{0.9838005039312139} \\
\textbf{Force}                & \textbf{\num{1.4627e-03}} / \textbf{\num[round-mode=places,round-precision=3]{0.9603730777104561}} &\num{1.7106e-03}  / \num[round-mode=places,round-precision=3]{0.9536571215855949} & \num{2.5762e-03} / \num[round-mode=places,round-precision=3]{0.9302063114897193} \\
\textbf{Combined}            & \textbf{\num{6.6127e-04}} / \textbf{\num[round-mode=places,round-precision=3]{0.9832341263637594}}  & \num{7.5532e-04}/ \num[round-mode=places,round-precision=3]{0.9806962809820803} & \num{1.2385e-03} / \num[round-mode=places,round-precision=3]{0.9687792791291034} \\ \bottomrule
\end{tabular}
\end{table}

\begin{table}
\caption{Accuracy metrics of FEMIN online runs compared to the BI load case test set with $\vec p = \vec 0$. Values are presented as MSE / $R^2$ score. Results of the LSTM networks are from \citet{thel_introducing_2024}.}
\label{tab:bi-np-test-performance}
\centering
\begin{tabular}{@{}lcccc@{}}
\toprule
\multicolumn{1}{c}{\textbf{QoI}} & \textbf{DVBF} & \textbf{LSTM full-seq.} & \textbf{LSTM window}  \\ \midrule
\textbf{Displacement} & \textbf{\num{7.8669e-05}} / \textbf{\num[round-mode=places,round-precision=3]{0.9980586747245861}} & \num{1.2545e-03} / \num[round-mode=places,round-precision=3]{0.969041582967119} & \num{2.4560e-04} /  \num[round-mode=places,round-precision=3]{0.9939394254954818} \\
\textbf{Velocity}& \textbf{\num{3.1416e-04}} / \textbf{\num[round-mode=places,round-precision=3]{0.9938582600051119} }& \num{2.5936e-03} /  \num[round-mode=places,round-precision=3]{0.9492966691200216}  & \num{8.0329e-04} / \num[round-mode=places,round-precision=3]{0.9842959839397644} \\
\textbf{Force } & \textbf{\num{1.5697e-03}} / \textbf{\num[round-mode=places,round-precision=3]{0.9574740488394327} }&  \num{2.0337e-03} / \num[round-mode=places,round-precision=3]{0.9449031638452157}& \num{2.6811e-03} /  \num[round-mode=places,round-precision=3]{0.9273635025432992} \\
\textbf{Combined} & \textbf{\num{6.5418e-04}} / \textbf{\num[round-mode=places,round-precision=3]{0.9831303278563769}} & \num{1.9606e-03} / \num[round-mode=places,round-precision=3]{0.9544138053107853}& \num{1.2433e-03} / \num[round-mode=places,round-precision=3]{0.968467} \\ \bottomrule
\end{tabular}
\end{table}

\Cref{tab:bi-test-performance} shows the accuracy performance of the DVBF compared to the deterministic LSTM approaches.
The DBVF significantly outperforms the window-based LSTM in all performance metrics despite being trained on a smaller window ($T^*=300$  vs. $T^*=600$).
Furthermore, the DBVF even surpasses the full-sequence LSTM (trained with backpropagation through the entire time sequence) in overall accuracy.
This is particularly noteworthy as the full-sequence LSTM, while highly accurate, lacks scalability for load cases with a large number of time steps.

The DVBF tends to produce smooth, rounded force prediction.
For example, in \cref{fig:bi-femin-curves}, the force prediction at $x=\SI{10}{\mm}$ in the x-direction (purple curve) does not follow every small swerving of the true data while still following the general force trajectory.  
The curve seems like a smoothed version of the ground truth.
This characteristic leads to a smaller $R^2$ value for the force prediction despite achieving a lower MSE.
This smoothing effect is a characteristic of variational autoencoders \citep{bredell_explicitly_2023} and does not compromise the DVBF's ability to represent the crash loads in this fast impact load case.
Instead, it seems that the DVBF corrects prediction error through the smoothing of the curve:
Analyzing again the x-force trajectory at $x=\SI{10}{\mm}$  at $2500 < t < 3500$, the DVBF first underestimates then overestimates the force, correcting the first prediction error. 
This error correction results in excellent displacement trajectories. 
Thus, the smoothing can be seen as a strength rather than a limitation.

The results in the configuration without design parameters ($\vec p = \vec 0$) are very similar. 
The predicted force trajectories have the same smoothed characteristics (see \cref{fig:bi-np-femin-curves}).
The accuracy metrics (\cref{tab:bi-np-test-performance}) show even slightly improved performance compared to the configuration with design parameters.
In this configuration ($\vec p = \vec 0$), the DVBF \textit{significantly} outperforms both full-sequence and window-based LSTM. 
This is despite the weights of the DVBF being chosen based on the lowest online NLL rather than the lowest online MSE. 
This suggests that the accuracy difference could potentially be even more significant if optimized solely for online accuracy, as it was done for the LSTM architectures.

Comparing the results in \cref{tab:bi-test-performance,tab:bi-np-test-performance}, it is apparent that the window-based LSTM and the DVBF (also windows-based training) do not show any performance degradation from excluding the design variables.
The initial network has to infer the latent states at the beginning of each window. 
These latent states are not constant. 
Therefore, the initial network has to rely on the kinematic (and force in the case of the DVBF) values. 
The DVBF also uses the design parameter vector $\vec p$ in the transition model (see \cref{equ:alpha-network,equ:locally-linear-transition}).
This implies that the encoder can infer these design parameters from the kinematic, so the vector $\vec p$ does not offer any new information.
The full-sequence LSTM's improvements from using the design parameters are likely due to its initial network, which only needs to predict states for $t=0$, where design parameters provide the best available information.
Nevertheless, these findings reinforce the general FEMIN approach of inferring forces based on kinematics.

\begin{figure}
    \centering
    \begin{subfigure}[t]{0.49\textwidth}
        \includegraphics[width=\textwidth]{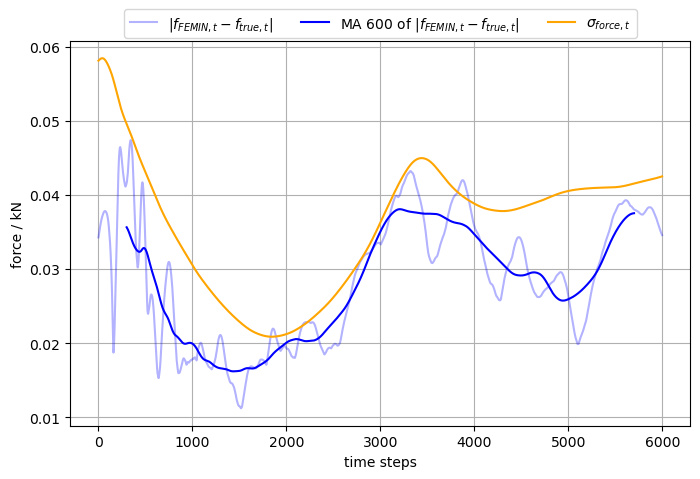}  
        \caption{$\vec p = \{\vec{v}_{\text{init}}, \vec{d}_{\text{impactor}}\}$}
    \end{subfigure}
    \hfill
    \begin{subfigure}[t]{0.49\textwidth}
        \includegraphics[width=\textwidth]{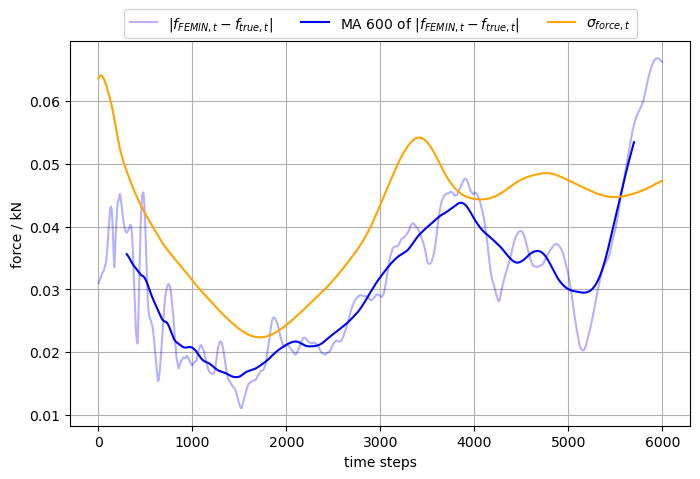}  
        \caption{$\vec p = \vec 0$}
    \end{subfigure}
    \caption{Comparison of the predicted standard deviation of the force output during the online application and the absolute difference of the predicted online force $\vec{f}_{\text{B}}$ to the true values. Results of a test sample from the BI load case are shown. The moving average (MA) is computed with a time window of 600. }
    \label{fig:bi-correlation}
\end{figure}

\begin{table}
\caption{Confidence metrics of BI load case with  $\vec p = \{\vec{v}_{\text{init}}, \vec{d}_{\text{impactor}}\}$ during the online application of the test samples.}
\label{tab:bi-confidence}
\centering
\begin{tabular}{@{}lccc@{}}
\toprule
\multicolumn{1}{c}{\textbf{QoI}} & PICP & AC & NLL  \\ \midrule
\textbf{Displacement} & \SI{88.33}{\percent} & \SI{1.5479}{\mm} & \num{-203.7397} \\
\textbf{Velocity}&  \SI{79.33}{\percent} &  \SI{1.3213}{\m\per\s} & \num{-94.6209} \\
\textbf{Force } &  \SI{88.23}{\percent} &  \SI{5.8506}{\kilo\newton} & \num{-156.0625}\\
\textbf{Combined} &  \SI{85.29}{\percent} &  - & \num{-454.4231} \\ \bottomrule
\end{tabular}
\end{table}

\begin{table}
\caption{Confidence metrics of BI load case with $\vec p = \vec 0$ during the online application of the test samples. }
\label{tab:bi-np-confidence}
\centering
\begin{tabular}{@{}lccc@{}}
\toprule
\multicolumn{1}{c}{\textbf{QoI}} & \textbf{PICP} & \textbf{AC} & \textbf{NLL}  \\ \midrule
\textbf{Displacement} & \SI{89.04}{\percent} & \SI{1.6314}{\mm} & \num{-209.3473} \\
\textbf{Velocity}&  \SI{77.03}{\percent} &  \SI{1.3121}{\m\per\s} & \num{-132.5401} \\
\textbf{Force } &  \SI{91.14}{\percent} &  \SI{6.6127}{\kilo\newton} & \num{-157.9411}\\
\textbf{Combined} &  \SI{85.74}{\percent} &  - & \num{-499.8284} \\ \bottomrule
\end{tabular}
\end{table}

The confidence metrics for the BI load case, presented in Tables \ref{tab:bi-confidence} and \ref{tab:bi-np-confidence}, demonstrate similar results for both parameterized and non-parameterized configurations.
Our analysis of confidence metrics focuses primarily on force predictions, as force is the only free variable in the FEMIN framework. 
The displacement and velocity errors are constrained because they will start at zero and tend to increase from there. 
In FEMIN, then NNs are trained offline with data extracted from previously run FEM simulations.
There is no coupling between NN and FEM during training, and the NN cannot learn how the specimen reacts to its force predictions. 
Consequently, the uncertainty provided by the decoder may not necessarily reflect the expected uncertainty in the kinematics of the online application.

The PICP for force predictions is high, with \SI{88.23}{\percent} ($\vec p = \{\vec{v}_{\text{init}}, \vec{d}_{\text{impactor}}\}$) and \SI{91.08}{\percent} ($\vec p = \vec 0$) of true values falling within the \SI{95}{\percent} confidence. 
This aligns well with theoretical expectations, indicating that the uncertainty estimation of the DVBF generalizes well to the online application.
The uncertainty estimation can also be seen in \cref{fig:bi-femin-curves,fig:bi-np-femin-curves}.
In general, the predicted uncertainty of the force increases in areas where the predictions differ from the true values, especially in the regions where the force prediction smooths out small oscillations. 
Contrary to the high PICP, the AC value is relatively small (\SI{5.8506}{\kilo\newton} and \SI{6.35}{\kilo\newton}) and only accounts for approximately \SI{10}{\percent} of the force trajectories range (\SI{65}{\kilo\newton}).
The AC value provides a human-interpretable measure of confidence in the FEMIN simulation. 
This metric is particularly valuable as it can be calculated without true values, offering insight into prediction reliability during the online application of new, unseen designs.

To further assess the relationship between prediction uncertainty and accuracy, we analyzed the correlation between the force standard deviation and the absolute error during online application. 
\Cref{fig:bi-correlation} illustrates this relationship, showing that the absolute error (light blue) follows the trend of the standard deviation but with more noise. 
A moving average was applied to reduce these fluctuations.
The Pearson correlation coefficient was computed between the smoothed absolute error and the predicted force standard deviation. 
The resulting correlations of $r = \SI{60.5}{\percent}$ for $\vec p = \{\vec{v}_{\text{init}}, \vec{d}_{\text{impactor}}\}$ and $r=\SI{61.4}{\percent}$ for $\vec p = \vec 0$ (averaged over all test samples) further establish the standard deviation as a valuable metric for assessing uncertainty during FEMIN application.

In conclusion, the standard deviation provided by the DVBF proves to be a valuable tool for estimating confidence in FEMIN simulations. 
However, it is important to note some limitations: the standard deviation is a trained parameter and may not accurately reflect uncertainty when the model is applied outside its training domain.
For example, if a part buckles in a FEMIN simulation that did not buckle in the FEM simulation used for the train data, it is not guaranteed that the predicted uncertainty increases. 
A similar conclusion was made by \citep{deshpande_probabilistic_2022}.
Despite these limitations, the DVBF's ability to provide meaningful uncertainty estimates within the trained domain represents a significant advancement in applying machine learning techniques to crash simulations.

\subsubsection{Tension-Compression-Tension (TCT)}

\begin{figure}
    \centering
    \begin{subfigure}[t]{0.33\textwidth}
        \includegraphics[width=\textwidth]{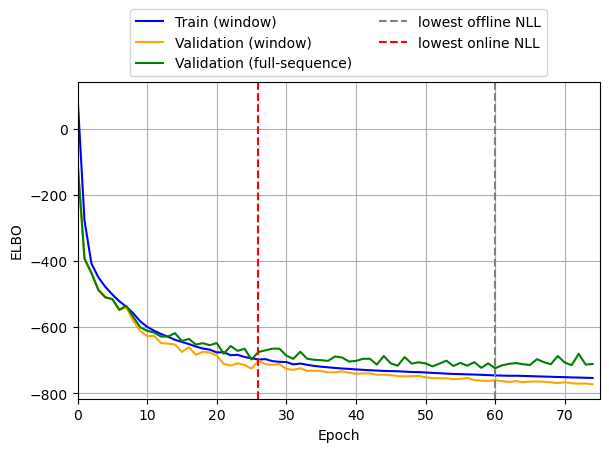}
        \label{fig:tct_elbo}
        \caption{ELBO (offline)}
    \end{subfigure}
    \hfill
    \begin{subfigure}[t]{0.33\textwidth}
        \includegraphics[width=\textwidth]{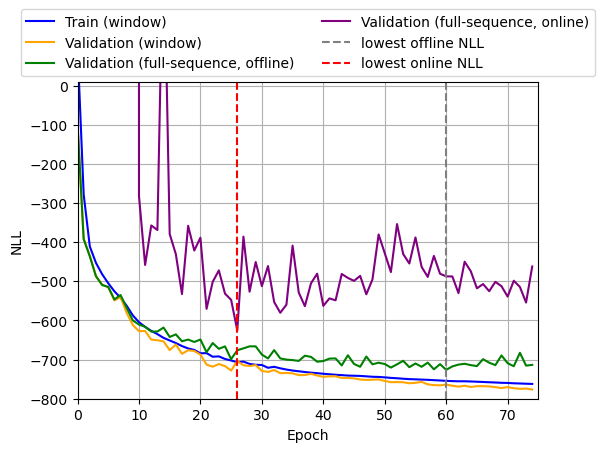}
        \label{fig:tct_nll}
        \caption{NLL (offline / online)}
    \end{subfigure}
    \hfill
    \begin{subfigure}[t]{0.33\textwidth}
        \includegraphics[width=\textwidth]{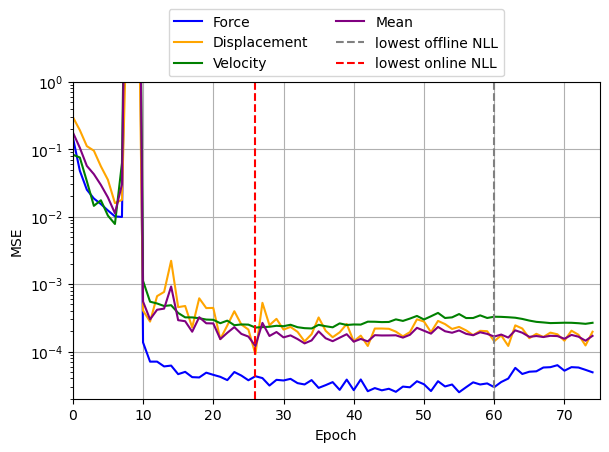}
        \label{fig:tct_mse}
        \caption{MSE (online)}
    \end{subfigure}
    \caption{Comparison of ELBO, NLL, and MSE over epochs for TCT load case}
    \label{fig:tct-loss}
\end{figure}

The training characteristics of the TCT load case, as illustrated in \cref{fig:tct-loss}, demonstrate a steady convergence pattern. 
Unlike the BI load case, there are no signs of overfitting, with train and validation losses remaining close together throughout the training process and the validation loss showing no increase. 
The DVBF exhibits excellent generalization from window-based training and validation to full-sequence validation, even in the early stages of training.
This is demonstrated by the close alignment of the window-based and full-sequence validation curves of the ELBO/NLL metric.
This suggests that the window-based approach effectively captures the history effects (plastic strain hardening) of the TCT scenario.
The online application is initially unstable, with the FEM failing to converge until around epoch 10. 
After this point, the online MSE converges along with the offline metrics. 
The online NLL shows more noise compared to its offline counterpart but does not diverge, indicating that the DVBF does not become overconfident during the training progress.

\begin{figure}
    \centering
    \includegraphics[width=\textwidth]{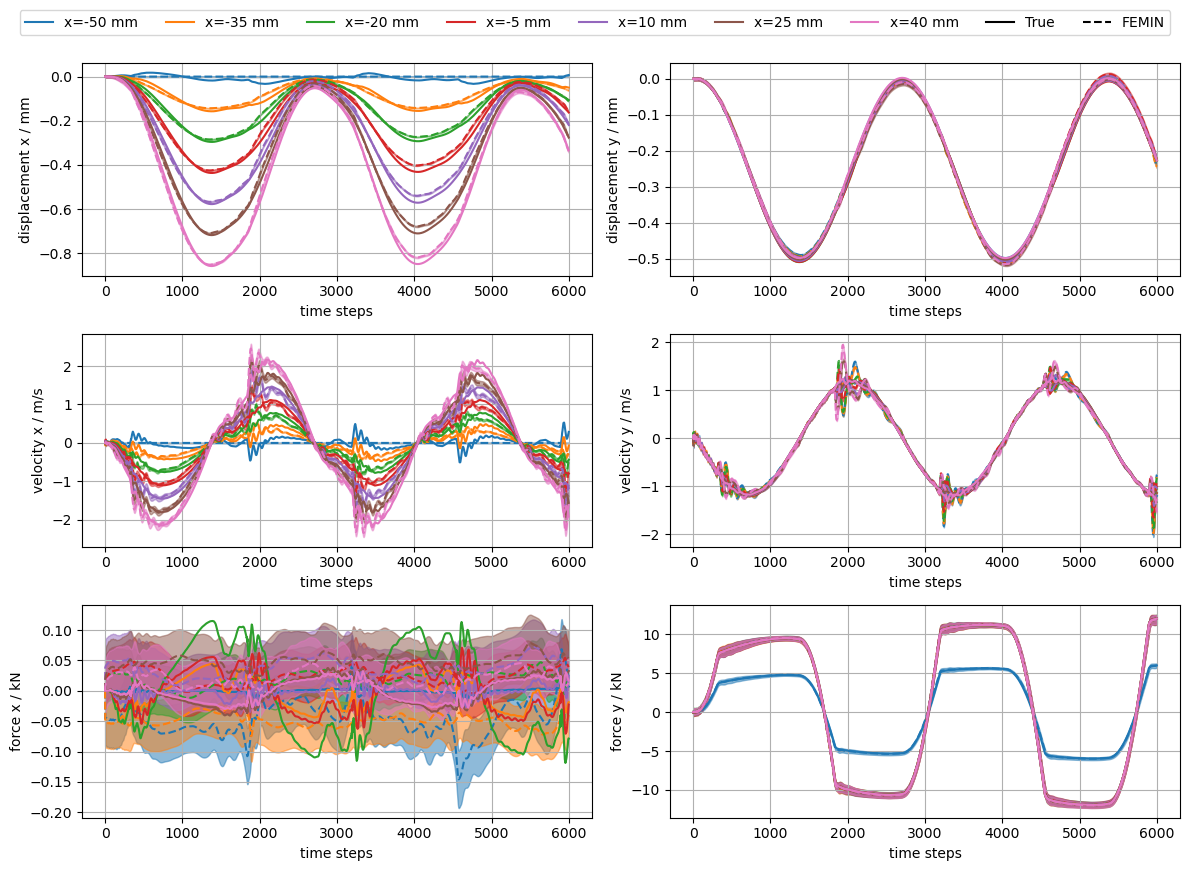}
    \caption{Online/FEMIN results of TCT load case for one test sample. The FEMIN run (dashed lines) is compared with the full FEM simulation (continuous lines). The results are given for selected nodes at the boundary $\mathcal{N}_{\text{B}}$. The shaded areas represent the standard deviation computed by the decoder.}
    \label{fig:tct-np-femin-curves}
\end{figure}

\begin{table}
\caption{Results of FEMIN online runs compared to the TCT load case test set. Values are presented as MSE / $R^2$ score. Results of the LSTM networks are from \citet{thel_introducing_2024}.}
\label{tab:tct-test-performance}
\centering
\begin{tabular}{@{}lccc@{}}
\toprule
\multicolumn{1}{c}{\textbf{QoI}} & \textbf{DVBF} & \multicolumn{1}{c}{\textbf{LSTM full-seq.}}      & \multicolumn{1}{c}{\textbf{LSTM window}}\\ \midrule
\textbf{Displacement}    &  \textbf{\num{3.5840e-04}} / \textbf{\num[round-mode=places,round-precision=3]{0.9972184669072919}}   & \num{9.1795e-03} / \num[round-mode=places,round-precision=3]{0.9287581434497822}  & \num{1.4721e-02} / \num[round-mode=places,round-precision=3]{0.8857502955110775} \\
\textbf{Velocity}          & \textbf{\num{8.2306e-04}}  / \textbf{\num[round-mode=places,round-precision=3]{0.9509569179284079} }    & \num{7.0184e-03} / \num[round-mode=places,round-precision=3]{0.5817600758496387} & \num{9.6284e-02} /\num[round-mode=places,round-precision=3]{-4.738106922246668} \\
\textbf{Force}              & \textbf{\num{8.1309e-05}}  / \textbf{\num[round-mode=places,round-precision=3]{0.9992837889607381}}  & \num{1.6373e-03}  / \num[round-mode=places,round-precision=3]{0.9855632416611118}  &   \num{2.5372e-03} /  \num[round-mode=places,round-precision=3]{0.9776267581024187}  \\
\textbf{Combined}           & \textbf{\num{4.2092e-04}} / \textbf{\num[round-mode=places,round-precision=3]{0.9824863912654793}}  & \num{5.9451e-03} / \num[round-mode=places,round-precision=3]{0.83202} & \num{3.7847e-02} / \num[round-mode=places,round-precision=3]{-0.95825}  \\ \bottomrule
\end{tabular}
\end{table}

The online application of the DVBF to the TCT load case demonstrates significant improvements over both LSTM models, as shown in Table \ref{tab:tct-test-performance}. 
The DVBF achieves MSE values of one order of magnitude better than the full-sequence LSTM and two orders of magnitude better than the window-based LSTM. 
Moreover, the DVBF exhibits substantially higher $R^2$ scores across all quantities. 
\Cref{fig:tct-np-femin-curves} illustrates that the DVBF accurately predicts the force resulting in kinematics that closely follow the true values from the full FEM simulation. 
Notably, vibration modes occur when the specimen reaches plasticity (see, for example, at $t=400$, $t=1800$, etc.).
These vibrations are most pronounced in the velocity signal due to their high frequency but low amplitude in the displacement signal.
The oscillations are most severe at the beginning of the plasticity and then gradually reduce. 
The FEMIN velocity signal settles to the true values within the plasticity.
Notably, during these vibrations, the predicted uncertainty of the force values increases, with the highest uncertainty occurring immediately after the kink in the force curve. 
The uncertainty then decreases while still in the plasticity phase, suggesting that the DVBF recognizes when the inputs deviate from expected patterns and adjusts its confidence accordingly. 
This behavior underscores the DVBF's ability to provide meaningful uncertainty estimates, particularly during complex material behavior transitions.

\begin{table}
\caption{Confidence metrics of TCT load case}
\label{tab:tct-confidence}
\centering
\begin{tabular}{@{}lccc@{}}
\toprule
\multicolumn{1}{c}{\textbf{QoI}} & \textbf{PICP} & \textbf{AC} & \textbf{NLL}  \\ \midrule
\textbf{Displacement} & \SI{66.99}{\percent} & \SI{0.01137}{\mm} & \num{17.18} \\
\textbf{Velocity}&  \SI{72.22}{\percent} &  \SI{0.09527}{\m\per\s} & \num{119.7692} \\
\textbf{Force } &  \SI{98.23}{\percent} &  \SI{0.6299}{\kilo\newton} & \num{-223.97}\\
\textbf{Combined} &  \SI{79.15}{\percent} &  - & \num{-100.02} \\ \bottomrule
\end{tabular}
\end{table}

\begin{figure}
    \centering
    \includegraphics[width=0.5\textwidth]{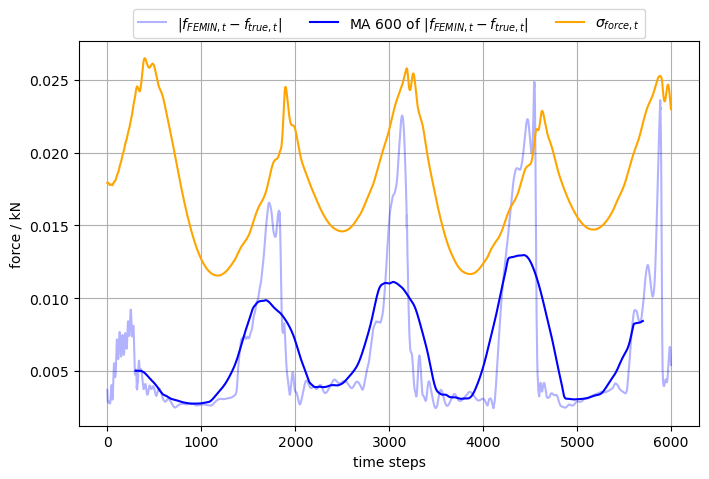}
    \caption{Comparison of the predicted standard deviation of the force output during the online application and the absolute difference of the predicted online force $\vec{f}_{\text{B}}$ to the true values. Results of a test sample from the TCT load case are shown. The moving average (MA) is computed with a time window of 600.}
    \label{fig:tct-correlation}
\end{figure}

The confidence metrics for the TCT load case are presented in \cref{tab:tct-confidence}.
The PICP for force predictions is exceptionally high at \SI{98.23}{\percent}, indicating that the \SI{95}{\percent} confidence interval effectively captures the differences between online predictions and true values. 
This suggests that the DVBF's uncertainty estimation for force is well-calibrated. 
The AC score for force is relatively small at \SI{0.6299}{\kilo\newton}, considering the force range of \SI{25}{\kilo\newton} in the simulation. 
This shows that the DVBF maintains high confidence in its force predictions throughout the simulation. 

\Cref{fig:tct-correlation} illustrates a clear correlation between the force prediction error and the predicted standard deviation over time.
Both metrics reach their respective maxima at the beginning of the plasticity region, coinciding with the previously observed vibrations in the velocity signal. 
This alignment suggests that the DVBF effectively recognizes increased uncertainty during complex material behavior transitions. 
The Pearson correlation coefficient between the smoothed absolute error and the predicted force standard deviation is $r = \SI{43.9}{\percent}$, lower than in the BI case. 
This lower correlation can be attributed to a slight lag between the standard deviation and the force error curves, as visible in \cref{fig:tct-correlation}.
Despite this lag, the overall trend shows that the DVBF's uncertainty estimates provide valuable insights into the reliability of its predictions.

\subsection{Computational efficiency}

\begin{table}
\caption{Training and inference time of different architectures. The training is reported as the time it takes the training to reach the epoch with the lowest loss value. The inference time is the time to compute the 6000 time steps sequentially. Results of the LSTM networks are from \citet{thel_introducing_2024}. All results are from an NVIDIA RTX A5000 using XLA \citep{sabne_xla_2020}.}
\label{tab:times}
\centering
\begin{tabular}{@{}lccc@{}}
\toprule
& \textbf{DVBF} & \textbf{LSTM full-seq.} & \textbf{LSTM window}  \\ \midrule
\textbf{Training BI} ($\vec p = \vec 0$) & \SI{37.68}{\minute}  & \SI{310.8}{\minute}  & \SI{8.4}{\minute}  \\
\textbf{Inference BI}  ($\vec p = \vec 0$)  &  \SI{4.65}{\s} & \SI{2.82}{\s} &\SI{4.95}{\s}  \\ \midrule
\textbf{Training TCT} & \SI{48.01}{\minute}  & \SI{192.8}{\minute}  & \SI{53.8}{\minute} \\
\textbf{Inference TCT}&  \SI{4.57}{\s} &  \SI{2.87}{\s} &  \SI{6.84}{\s}  \\
\end{tabular}
\end{table}

The computational efficiency of the DVBF architecture, compared to the LSTM models, is presented in Table \ref{tab:times}. 
The DVBF demonstrates competitive performance in both training and inference times. 
For training, the DVBF significantly outperforms the full-sequence LSTM, requiring only \SI{37.68}{\minute} for the BI load case compared to  \SI{310.8}{\minute} minutes for the LSTM. 
While the DVBF training time is longer than the window-based LSTM for the BI case due to the computational cost of variational inference, it shows superior training efficiency for the TCT load case.
Here, the DVBF trains in \SI{48.01}{\minute}, compared to \SI{53.8}{\minute} for the LSTM window approach. 
This performance gain is likely due to the window-based LSTM's difficulty in converging for the TCT scenario, requiring significantly more epochs than in the BI case.

Regarding inference times, the DVBF maintains competitive performance across both load cases. 
While slower than the full-sequence LSTM (\SI{4.65}{\s} vs \SI{2.82}{\s} for BI, \SI{4.57}{\s} vs \SI{2.87}{\s} for TCT), it consistently outperforms the window-based LSTM approach. 
However, it should be noted that \citet{thel_introducing_2024} reported that the difference in inference time between full-sequence and window-based LSTM was likely due to compilitation differences in XLA \citep{sabne_xla_2020}.
These results demonstrate that the DVBF is not only suitable for FEMIN in terms of accuracy and uncertainty estimation but also offers competitive training efficiency and inference speed. 
The DVBF's ability to provide probabilistic outputs without significantly compromising computational performance makes it a valuable tool for enhancing the reliability and efficiency of crash simulations in the FEMIN framework.

\section{Conclusions}

This study has demonstrated the successful adaptation of the Deep Variational Bayes Filter (DVBF) to the Finite Element Method Integrated Networks (FEMIN) framework, offering significant improvements in terms of accuracy while adding uncertainty estimation. 
Key findings include:
\begin{enumerate}
\item The DVBF consistently outperformed full-sequence and window-based LSTM architectures across multiple accuracy metrics.
\item The probabilistic nature of DVBF provided valuable uncertainty estimates, with high Prediction Interval Coverage Probability (PICP) and meaningful Average Confidence (AC) scores.
\item A strong correlation between predicted standard deviations and actual force prediction errors was observed, validating the DVBF's uncertainty estimates as a reliable prediction confidence indicator.
\item The DVBF demonstrated competitive computational efficiency in both training and inference.
\end{enumerate}
These results highlight the DVBF's potential to enhance the robustness and reliability of FEMIN simulations. 

Future work should focus on scaling the DVBF approach to larger, more complex load cases and exploring methods to improve the alignment between offline training metrics and online performance. 
Additionally, investigating the DVBF's behavior in scenarios outside its training domain could provide insights into its generalization capabilities and limitations.
In conclusion, the integration of DVBF into FEMIN represents a significant step forward in combining the efficiency of neural networks with the reliability requirements of crash simulations.

\appendix

\section{Hyperparameter} \label{sec:hyperparameter}

The Lion optimizer \citep{chen_symbolic_2023} is used for training the DVBF.
The learning rate is changed every mini-batch $b$ (batch size is 256) based on a linear warm-up and exponential decay:
\begin{equation}
\text{lr}(b) = 
\begin{cases} 
\beta \cdot \frac{b}{B_w} & \text{if } b < B_w \\
\beta \cdot \gamma^{\left(\frac{b - B_w}{B_d}\right)} & \text{if } b \geq B_w 
\end{cases}
\end{equation}
where\\[1.5ex]
\begin{tabular}{ll}
    $\beta$ & maximum learning rate \\
    $B_w$ & number of warm-up steps \\ 
   $B_d$ & number of decay steps\\
   $\gamma$ & decay rate\\
\end{tabular}
\\[1.5ex]
The decoupled weight decay is constant:
\begin{equation}
    \text{wd}(b) = \text{wd}
\end{equation}

\subsection{BI load case}

\begin{itemize}
    \item Latent dimension: $z \in \mathbb{R}^{32}$
    \item Time window length: $T^*=300$
    \item Optimizer: $\beta =\num{6.13e-5}$, $B_w=300$,  $B_d=\num{10000}$, $\gamma=0.64$,  $ \text{wd}=0.4$
    \item Gaussian noise added to input of $q_{\text{init}}$ and $q_{\text{enc}}$: $\sigma = 0.1$
    \item Initial network: $K=7$; MLP with $5$ layers á $256$ units with GeLU activation \citep{hendrycks_gaussian_2023}; output head for $\vec{\mu}_{w}$ with $32$ units and linear activation; output head for $\vec{\mu}_{w}$ with $32$ units and softplus activation
    \item Initial transition: MLP with $1$ layers á $64$ units with GeLU activation a output head with $32$ units with linear activation
    \item Encoder: MLP with $4$ layers á $256$ units with GeLU activation; output head for $\vec{\mu}_{\text{enc}}$ with $32$ units and linear activation; output head for $\vec{\mu}_{\text{enc}}$ with $32$ units and softplus activation 
    \item Transition: $M=128$, $s=0.2$, alpha network: MLP with $1$ layers á $64$ units with GeLU activation and a output head $32$ units and sigmoid activation 
    \item Decoder: MLP with $4$ layers á $256$ units with GeLU activation; output head for $\vec{\mu}_{\text{lik}}$ with $189$ units and linear activation; output head for $\vec{\mu}_{\text{lik}}$ with $189$ units and softplus activation
\end{itemize}
\subsection{TCT load case}

\begin{itemize}
    \item Latent dimension: $z \in \mathbb{R}^{64}$
    \item Time window length: $T^*=300$
    \item Optimizer: $\beta =\num{2.35e-5}$, $B_w=300$,  $B_d=\num{10000}$, $\gamma=0.7$,  $ \text{wd}=1$
    \item Gaussian noise added to input of $q_{\text{init}}$ and $q_{\text{enc}}$: $\sigma = 0.18$
    \item Initial network: $K=9$; MLP with $3$ layers á $128$ units with GeLU activation; output head for $\vec{\mu}_{w}$ with $64$ units and linear activation; output head for $\vec{\mu}_{w}$ with $64$ units and softplus activation
    \item Initial transition: MLP with $2$ layers á $128$ units with GeLU activation a output head with $64$ units with linear activation
    \item Encoder: MLP with $2$ layers á $512$ units with GeLU activation; output head for $\vec{\mu}_{\text{enc}}$ with $64$ units and linear activation; output head for $\vec{\mu}_{\text{enc}}$ with $64$ units and softplus activation 
    \item Transition: $M=16$, $s=0.2$, alpha network: MLP with $3$ layers á $256$ units with GeLU activation and a output head $64$ units and sigmoid activation 
    \item Decoder: MLP with $4$ layers á $512$ units with GeLU activation; output head for $\vec{\mu}_{\text{lik}}$ with $189$ units and linear activation; output head for $\vec{\mu}_{\text{lik}}$ with $189$ units and softplus activation
\end{itemize}

\newpage